\documentclass[aps,twocolumn,english,superscriptaddress,citeautoscript,preprintnumbers,amsmath,amssymb,floatfix,footinbib]{revtex4-2}
\usepackage[breaklinks=true,colorlinks,citecolor=blue,linkcolor=blue,urlcolor=blue]{hyperref}
\usepackage{amsmath}
\usepackage{graphicx}
\usepackage{dcolumn}
\usepackage{float}
\usepackage[utf8]{inputenc}
\DeclareUnicodeCharacter{2212}{-}
\usepackage{soul}

\usepackage{color}
\usepackage[super]{nth}

\usepackage{mathtools}
\DeclarePairedDelimiter\bra{\langle}{\rvert}
\DeclarePairedDelimiter\ket{\lvert}{\rangle}
\begin{document}
	
\title{Crystalline electric field and large anomalous Hall effect in the candidate topological material CeGaSi}
\author{Rajesh Swami}
\affiliation{Department of Physics, Indian Institute of Technology, Kanpur 208016, India}
\author{Daloo Ram}
\affiliation{Department of Physics, Indian Institute of Technology, Kanpur 208016, India}
\affiliation{School of Advanced Materials, and Chemistry and Physics of Materials Unit, Jawaharlal Nehru Centre for Advanced Scientific Research, Jakkur, Bangalore 560064, India}
\author{Anusree C.V.}
\affiliation{Department of Physics, Indian Institute of Technology Hyderabad, Kandi, Sangareddy 502 285, Telangana, India}
\author{V. Kanchana}
\email{kanchana@phy.iith.ac.in}
\affiliation{Department of Physics, Indian Institute of Technology Hyderabad, Kandi, Sangareddy 502 285, Telangana, India}
\author{Z. Hossain}
\email{zakir@iitk.ac.in}
\affiliation{Department of Physics, Indian Institute of Technology, Kanpur 208016, India}


\begin{abstract}	
We report a comprehensive investigation of CeGaSi single crystals, including magnetic, thermodynamic, electronic, and magnetotransport properties. The powder x-ray diffraction refinement revealed that CeGaSi crystallizes in LaPtSi-type tetragonal structure with space group $I$4$_1$\textit{md}. The electrical resistivity data show a metallic nature with a sharp drop occurring around $T_m \sim$ 11 K, revealing a magnetic phase transition, which is confirmed by magnetic susceptibility and heat capacity data. The magnetic susceptibility, magnetization, and heat capacity data are analyzed through the crystalline electric field based on point charge model, suggesting that the six degenerate ground states of Ce$^{3+}$ ($J$ = 5/2) ion split into three doublets with an overall splitting energy $\sim$ 288 K. The maximum negative magnetoresistance in CeGaSi for both $B$$\parallel$$c$ and $B$$\parallel$$ab$ field-direction is observed near $T_m$, it is attributed to the suppression of spin-disorder scattering by the magnetic field. The Hall resistivity data for $B$$\parallel$$c$ and $B$$\parallel$$ab$ show anomalous Hall signal. Our scaling analysis suggests that anomalous Hall effect in CeGaSi is dominated by the skew scattering mechanism. In addition, first-principles calculations identify CeGaSi as a nodal-line metal.
\end{abstract}

\maketitle
\section{Introduction}	
The cerium-based equiatomic intermetallic compounds CeTX (T = transition metals and X = p-block elements) have drawn great attention due to their interesting experimental findings, such as long-range ordering, valence fluctuations, strong crystalline electric field (CEF) effect, and various topological electronic phases \cite{CeRhSn,RGaGe_daloo,NdGaGe_daloo,CeNiSn,CeNiSn_theory}. The realization of a Weyl semimetallic state in these materials is especially notable, driven by the breaking of time-reversal symmetry (TRS) due to their spontaneous magnetic ordering. The interplay between Weyl-fermions and magnetism emerges various anomalous transport properties in these materials \cite{CeAlSi_Hall02,CeAlSi_loop_Hall_effect,CeAlGe_Hall,CeAlGe_THE,CeAlSi_2021,CeAlSi_2023,CeGaSi_2024}. Additionally, the presence of CEF potential can be split the ground state of Ce$^{3+}$ ions in compounds, resulting in the temperature-dependent thermodynamic and transport properties of these compounds are hugely affected \cite{RGaGe_daloo,Ce2Pd2Pb_CEF,CeAgAs2_2018}. 

Recently, the CeXY (X = Al/Ga and Y = Si/Ge) compounds have attracted significant attention due to the presence of a Weyl semimetallic state. This state arises from the breaking of both inversion symmetry (IS) and TRS, as these compounds crystallize in a noncentrosymmetric LaPtSi-type structure with space group $I$4$_1$\textit{md} (No. 109) and long-range magnetic ordering, respectively \cite{CeAlSi_2021,RAlGe_2018_theory,RAlGe_2019,CeGaSi_2024,CeAlGe_2018,RGaGe_daloo}. The position of Weyl nodes in these materials is easily tuned by magnetization, which induces to the nonzero Berry curvature, resulting in intriguing transport properties. For instance, CeAlGe displays the topological Hall effect, while CeAlSi shows an intrinsic anomalous Hall effect (AHE) when a magnetic field is applied along the easy magnetic axis and exhibits a loop Hall effect when the field is aligned with the hard magnetic axis \cite{CeAlGe_THE,CeAlSi_2021}. Among them, the CeGa$_x$Si$_{2-x}$ alloy is an interesting material, characterized by valence fluctuation with a lack of magnetic ordering in its parent compound CeSi$_2$ down to 100 mK \cite{CeSi2_1982}. This nonmagnetic ground state persists within the composition range 0 $\leq$ $\textit{x}$ $<$ 0.2. In the range 0.7 $\leq$ $\textit{x}$ $\leq$ 1.4, it shows ferromagnetic (FM) ordering around 10 K, while for $\textit{x}$ $\geq$ 1.5, it orders antiferromagnetically \cite{CeSi2-xGa_1984,CeSi2-xGa_1990, CeSi2-xGa_2022,CeSi2-xGa_1993}. It is also noteworthy to mention that the Ga concentration affects the electrical resistivity nature of CeGa$_x$Si$_{2-x}$ compound. For example, at $x$ = 0.7, its electrical resistivity demonstrates semiconductor-like behavior, while it exhibits metallic behavior at $x$ = 0.8, 1 \cite{CeSi2-xGa_1998,CeSi2-xGa_2022}. These findings contradict recent studies on CeGaSi single crystals that report a semiconductor-like behavior in electrical resistivity \cite{CeGaSi_2024, CeGaSi_2024_02}, indicating that the physical properties of CeGaSi have strong sensitivity to actual chemical composition and the structural parameters. Our system, CeGa$_x$Si$_{2-x}$ ($x$ = 1), becomes an interesting candidate to explore the magnetic and transport properties in its metallic state.

In this study, we investigate the physical properties of CeGaSi single crystals, which are synthesized using arc-melted polycrystalline CeGaSi and gallium flux. Powder x-ray diffraction (XRD) refinement reveals that the compound CeGaSi crystallizes in a noncentrosymmetric tetragonal structure with the space group $I$4$_1$\textit{md}. Our magnetic susceptibility, heat capacity, and electrical resistivity data suggest that CeGaSi undergoes magnetic ordering at $T_m$ $\thicksim$ 11 K. CEF analysis of magnetic susceptibility, magnetization, and heat capacity data indicates that the $J$ = 5/2 multiplet of Ce$^{3+}$ splits into three doublets. The temperature-dependent electrical resistivity of CeGaSi exhibits metallic behavior. In the ordered state, the magnetoresistance (MR) for $B$$\parallel$$ab$ shows a crossover from negative at low magnetic fields to positive at higher fields. Furthermore, we have observed a large AHE in- and out-plane Hall resistivity data. A large AHE in CeGaSi is explained through the skew scattering mechanism as confirmed by our scaling analysis. Furthermore, this study incorporates first-principles calculations to explore the magnetic ground state, electronic structure, and topological properties of CeGaSi. These key findings give valuable insights into the electrical and physical properties of CeGaSi single crystal and also distinguish it from recent studied by Gong et al. \cite{CeGaSi_2024} and Zhang et al. \cite{CeGaSi_2024_02}.


\section{Methods}
\label{II}
\begin{figure}
	\includegraphics[width=7.2cm, keepaspectratio]{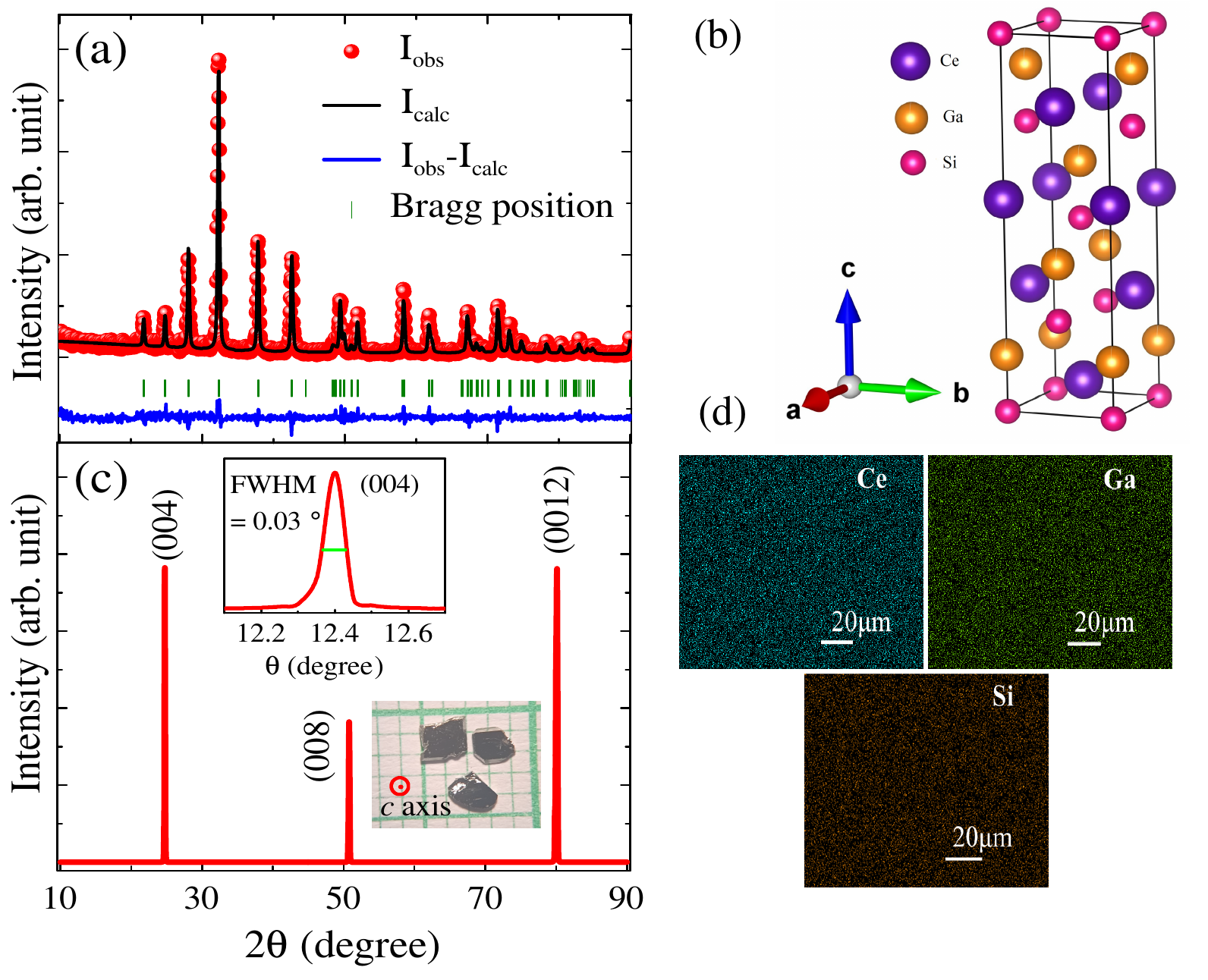}
	\caption{\label{Fig1}(a) Powder XRD pattern of crushed single crystals of CeGaSi recorded at room temperature, showing experimental data (red circles) and calculated data using the Rietveld refinement (black line). The blue line shows the difference between experimental and calculated data. Olive tick marks represent Bragg positions. (b) The crystal structure of CeGaSi for space group $I$4$_1$\textit{md}. (c) XRD pattern on single crystal of of CeGaSi. Lower and upper insets display the photograph of single crystals and rocking curve, respectively. (d) Elemental mapping analysis of CeGaSi using EDS.} 
\end{figure}

Single crystals of CeGaSi were grown by gallium flux and a precursor polycrystalline CeGaSi sample. The polycrystalline CeGaSi was synthesized using the standard arc melting technique. The starting elements of Ce pieces (99.9\%, Alfa Aesar), Ga ingot (99.999\%, Alfa Aesar), and Si pieces (99.9999\%, Alfa Aesar) are taken in stoichiometric ratio and put on water-cooled copper hearth under an argon atmosphere. After each melting, the sample was flipped and melted several times to ensure homogeneity. Then, polycrystalline CeGaSi and Ga were weighted in a 1:7 molar ratio and loaded into an alumina crucible. Next, the crucible was sealed in a quartz tube under partial argon pressure. The ampule was placed in the furnace and heated to 1100 $^\circ$C at a rate of 50 $^\circ$C/h, held for 10 h. Then, the furnace was slowly cooled down to 500 $^\circ$C at a rate of 3 $^\circ$C/h. At this stage, it was quickly taken out from the furnace and centrifuged to remove the excess Ga flux. This process yielded plate-like shiny single crystals with typical dimensions of \mbox{2 $\times$ 2 $\times$ 0.3 mm$^3$}, as shown in the lower inset of Fig. \ref{Fig1}(c).

The crystal structure and orientations were confirmed through the XRD using a PANalytical X’Pert PRO diffractometer with Cu K$_{\alpha1}$ radiation. The elemental analysis of single crystals was investigated by energy-dispersive x-ray spectroscopy (EDS) in a JEOL JSM-6010LA electron microscope. The magnetic susceptibility and magnetization measurements were performed by Physical Property Measurement System (PPMS) Quantum Design using the vibrating sample magnetometer. Heat capacity was measured using the relaxation method in the same PPMS. The magnetotransport measurements were carried out by the standard four-probe method in the ac mode using the Cryogen Free measurement system, Cryogenic Ltd., with a magnetic field 12 T.

The first-principles calculations were performed using the highly accurate Projector Augmented Wave (PAW) method \cite{paw}, as implemented in the \emph{Vienna Ab Initio Simulation Package} (\textsc{VASP}) \cite{vasp_1,vasp_2}. A robust plane-wave energy cutoff of 600~eV was adopted for the basis set, ensuring high precision. The exchange-correlation effects were treated within the framework of the Generalized Gradient Approximation (GGA), specifically employing the Perdew–Burke–Ernzerhof (PBE) parametrization \cite{pbe}. To sample the irreducible Brillouin zone (BZ), a dense $\Gamma$-centered $k$-point mesh of $12 \times 12 \times 6$ was utilized. The ground-state energy was determined by relaxing both atomic positions and lattice parameters using the conjugate-gradient algorithm, with convergence criteria set to an atomic force tolerance of $10^{-2}~\text{eV}/\text{\AA}$ and a total energy tolerance of $10^{-8}~\text{eV}$. To accurately capture the behavior of the Ce-$f$ electrons, the GGA+U method, as formulated by Dudarev \textit{et al.}~\cite{dudarev.botton.98}, was employed. An effective Hubbard parameter, $U_\text{eff} = U - J$ (with $J = 0$), was chosen with $U_\text{eff}$ set to 7~eV, enabling precise magnetic and electronic structure calculations \cite{CeAlSi_Hall02}.

\section{Results and discussion }
\label{III}

\subsection{Crystal structure}
Figure \ref{Fig1}(a) shows the powder XRD pattern of crushed CeGaSi single crystals at recorded room temperature. Further, we have analyzed the powder XRD data using the Rietveld refinement method in the FULLPROF software. The refinement reveals that compound CeGaSi crystallizes in a tetragonal structure with space group $I$4$_1$\textit{md}. The obtained lattice parameters are \textit{a} = \textit{b} = 4.2452 \AA{}, and \textit{c} = 14.3443 \AA{}. The structure arrangement of CeGaSi contains four formula units per unit cell, as shown in Fig. \ref{Fig1}(b). All three constituent atoms (Ce, Ga, and Si) occupy the 4\textit{a} Wyckoff site. Furthermore, there exists a stacking arrangement of Ce, Ga, and Si layers in an alternative manner along the \textit{c} axis with a lack of IS. The single crystal XRD pattern of CeGaSi is measured on a flat surface of the rectangular-shaped single crystal, as presented in Fig. \ref{Fig1}(c). The presence of (00$l$) peaks indicates that the \textit{c} axis is perpendicular to the crystal surface. These peaks are very sharp, as evidenced by a small full width at half maximum (FWHM, $\Delta$$\theta$ = 0.03$^\circ$) of the rocking curve around the (004) peak, as depicted in the upper inset of Fig. \ref{Fig1}(c), confirming the grown crystals in good quality. The EDS results suggest that all elements (Ce, Ga, and Si) are evenly distributed with expected equiatomic stoichiometry in the sample.

\subsection{Magnetic properties}

\begin{figure}
	\centering
	\includegraphics[width=8.6cm, keepaspectratio]{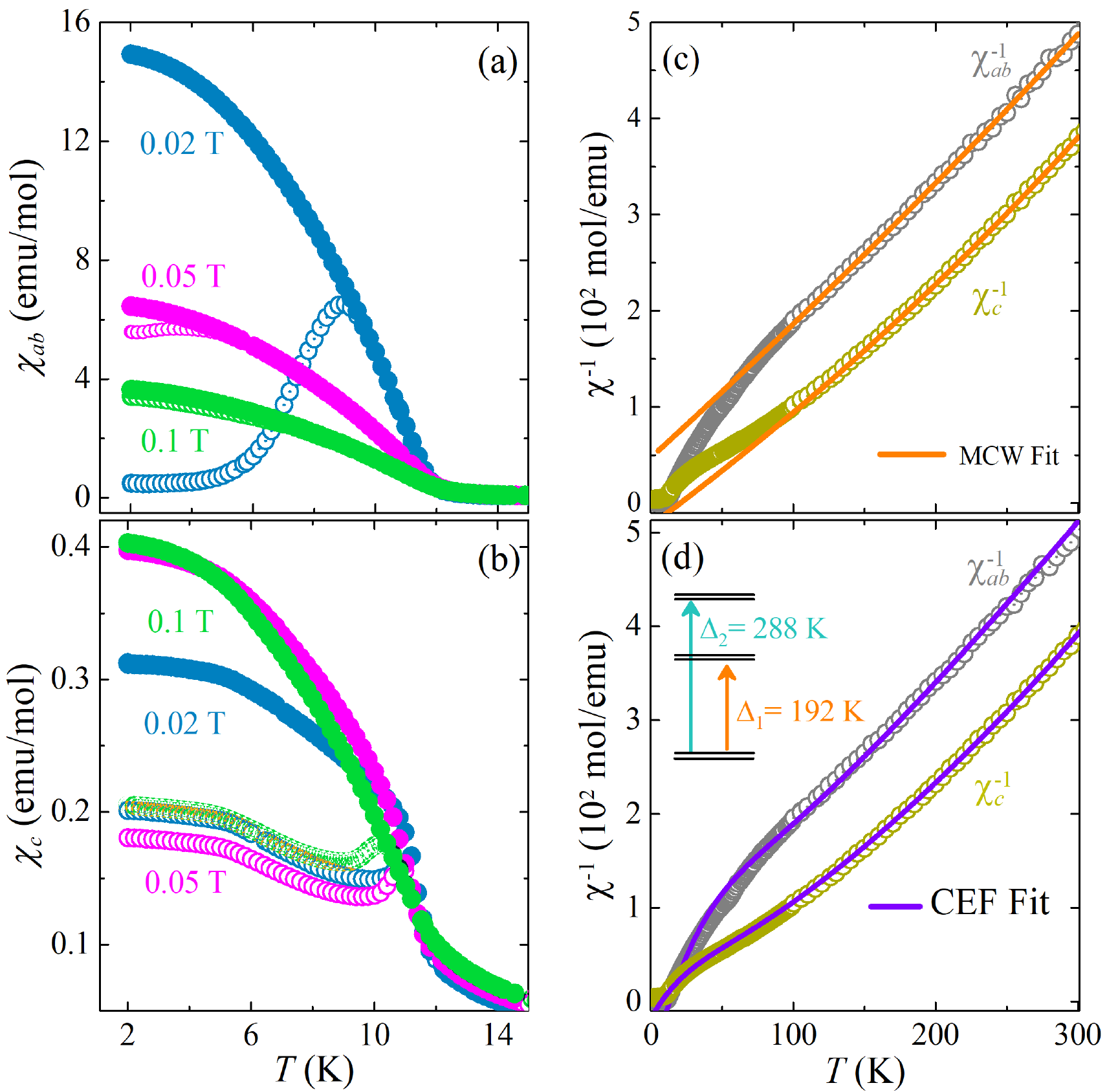}
	\caption{\label{Fig2}The temperature dependence of magnetic susceptibility $\chi$ for (a) $B$$\parallel$$ab$ ($\chi$$_{\textit{ab}}$) and (b) $B$$\parallel$$c$ ($\chi$$_{\textit{c}}$) under various applied fields. Open and filled circles represent ZFC and FC configurations, respectively. (c) and (d) The inverse magnetic susceptibilities as a function of temperature. Solid orange lines present the MCW fit in (c). The solid violet lines show the fit obtained using Eq. \ref{CEF1} in (d).}
\end{figure}

Figures \ref{Fig2}(a) and \ref{Fig2}(b) illustrate the temperature dependence of magnetic susceptibility $\chi$ along the crystallographic \textit{ab} and \textit{c} ($\chi$$_\textit{ab}$ and $\chi$$_\textit{c}$) directions in the zero field-cooled (ZFC) and field-cooled (FC) modes under different applied magnetic fields. At $B$ = 0.02 T, the ZFC $\chi$$_\textit{ab}$ and $\chi$$_\textit{c}$ data show a clear peak at $T_m$ ($\sim$ 11 K) due to the magnetic ordering. After a sharp drop, $\chi$$_\textit{ab}$ and $\chi$$_\textit{c}$ starts to increase again below 3.0 and 9.5 K, respectively, signaling presence of a weak FM component \cite{TbNi2B2C}. Below $T_m$, both FC curves show a pronounced increase with decreasing temperature and gradually tend to saturate at lower temperatures. Additionally, $\chi$($T$) shows a pronounced bifurcation between ZFC and FC data below $T_m$. With increasing field, the peak at $T_m$ shifts to lower temperatures, a characteristic signature of an AFM system \cite{GdAuGe}. At $B$ = 0.1 T for $B$$\parallel$$ab$, the peak is suppressed, and both ZFC and FC curves start to overlap and become almost flat, indicative of a field-induced FM state in CeGaSi. Interestingly, $\chi$$_\textit{ab}$ is approximately 40 times larger than $\chi$$_\textit{c}$ at 2 K, indicating a strong magnetic anisotropy in CeGaSi single crystals. 
Next, the inverse magnetic susceptibility data above 150 K are fitted using the modified Curie-Weiss law (MCW), $\chi$(\textit{T}) = $\chi_0$ + $C$/(\textit{T} - $\Theta_{CW}$), where \textit{C} is the Curie constant, $\Theta_{CW}$ is the Curie-Weiss temperature, and $\chi_0$ is temperature independent magnetic susceptibility, as presented in the Fig. \ref{Fig2}(c). However, below 150 K, the $\chi^{-1}$($T$) deviates from the MCW law and exhibits a hump around 70 K due to the presence of crystal field potential of Ce$^{3+}$ ion, as observed in various rare-earth compounds such as CeAg$X$$_2$ (X = As, Sb) \cite{CeAgAs2_2018,CeAgSb2_2005} and $R$GaGe ($R$ = Ce-Nd) \cite{RGaGe_daloo,NdGaGe_daloo}. The obtaining fitting parameters are $\chi_0$ = -1.98 (-5.18) $\times$10$^{-4}$ emu/mol and $\Theta_{CW}^{ab}$ = -36.1 ($\Theta_{CW}^{c}$ = 20.9) K for \textit{B}$\parallel$\textit{ab} (\textit{B}$\parallel$\textit{c}). The opposite sign of $\Theta_{CW}$ along two directions indicate that the FM interactions are dominant along the $c$ axis, while AFM interactions are dominant in $ab$ plane. The calculated effective magnetic moment $\mu_{eff}$ = 2.43 $\mu_B$ (for \textit{B}$\parallel$\textit{ab}) and  2.62 $\mu_B$ (for \textit{B}$\parallel$\textit{c}) are close to expected value 2.54 $\mu_B$ for free Ce$^{3+}$ ion. A similar temperature dependence of magnetic behavior has also been observed in isostructural compounds such as CeGaGe and CeAlSi \cite{CeAlSi_2021,RGaGe_daloo}. 
\begin{figure*}
	\includegraphics[width=17cm, keepaspectratio]{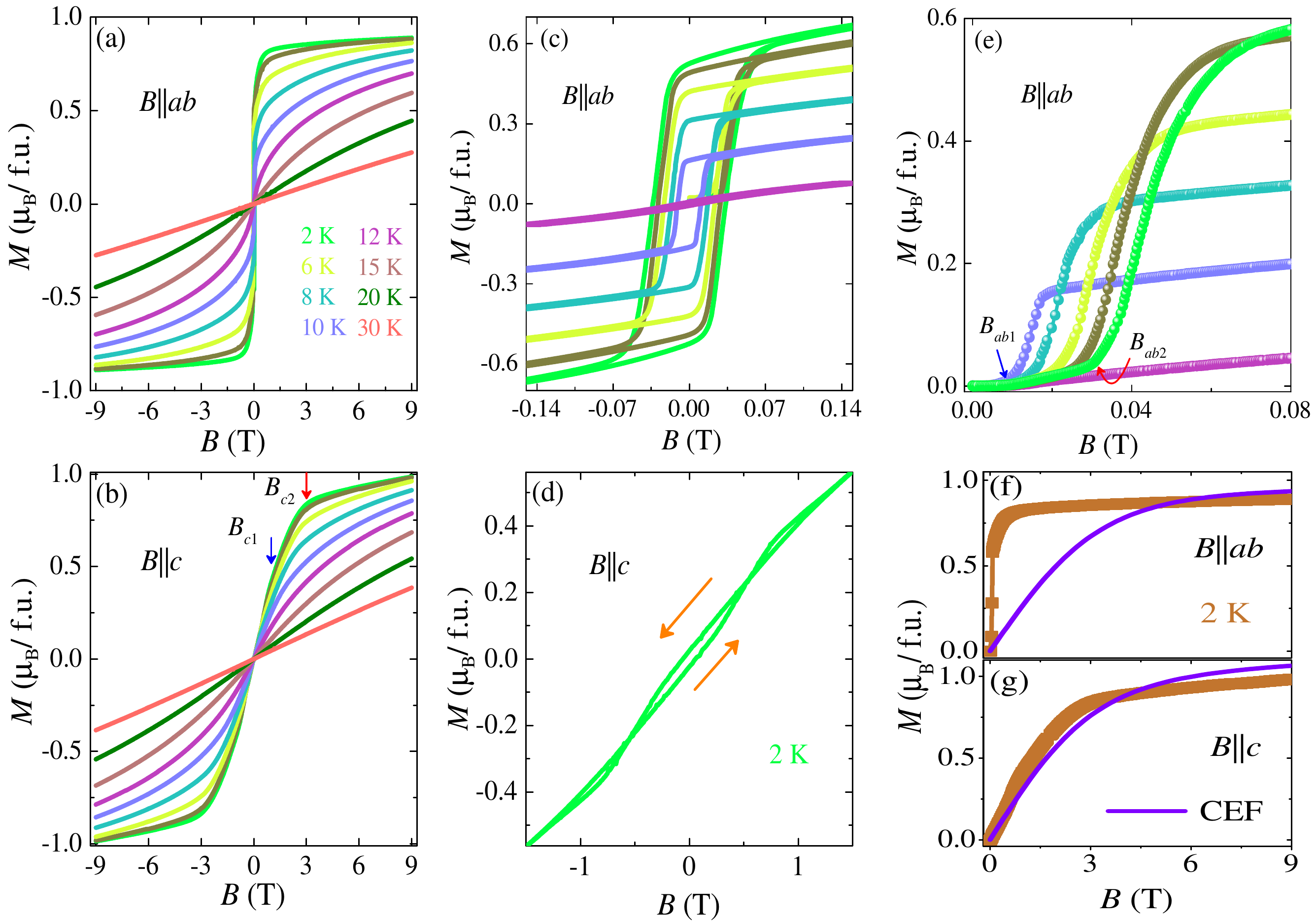}
	\caption{\label{Fig3}Isothermal magnetization $M$($B$) at different temperatures for (a) $B$$\parallel$$ab$ and (b) $B$$\parallel$$c$. $M$($B$) loops at selected temperature for (c) $B$$\parallel$$ab$ and (d) $B$$\parallel$$c$. (e) The low-field virgin magnetization curves at different temperatures for $B$$\parallel$$ab$}. $M$($B$) at 2 K for (f) $B$$\parallel$$ab$ and (g) $B$$\parallel$$c$, along with the calculated magnetization from the CEF scheme shown as violet solid lines.
\end{figure*}

Further, we have analyzed the magnetic susceptibility data using CEF scheme based on the point charge model. Ce$^{3+}$ ions in CeGaSi have a tetragonal site symmetry and C$_{4v}$ point symmetry, then six ground states ($J$ = 5/2 multiplet) of Ce atoms split into three doublets. In tetragonal site symmetry of Ce ions, we used the following CEF Hamiltonian.
\begin{equation}
	\mathcal{H}_{CEF} = B_2^0O_2^0+B_4^0O_4^0+B_4^4O_4^4,
	\label{CEF}
\end{equation}
here, $B_l^m$ and $O_l^m$ denote the CEF parameters and the Stevens operators, respectively \cite{CEF_1951,CEF_1964}.
\noindent The overall inverse magnetic susceptibility, including the molecular field contribution $\lambda_i$ and the residual susceptibility $\chi_0^i$, is given by
\begin{equation}
	\label{CEF1}
	\chi_i^{-1}
	=\bigg(\frac{\chi_{CEF}^i}{1-\lambda_i\chi_{CEF}^i}
	+\chi_0^i\bigg)^{-1},
\end{equation}
where $\chi_{CEF}^i$ is the temperature dependence of the susceptibility based on CEF model as expressed in Refs. \cite{RGaGe_daloo,CeAgSb2_2005}. The computed inverse magnetic susceptibility is well-fitted to the experimental data for both field orientations, as shown in Fig. \ref{Fig2}(d). The estimated values of CEF parameters, energy levels, and $\chi_0^i$ of CeGaSi are listed in Table \ref{CEF_chi}. The $\chi_0^i$ values are consistent with obtained from the fitting of $\chi^{-1}(T)$ using MCW law. Moreover, we have also estimated the first CEF parameter $B_2^0$ directly from the Curie-Weiss temperatures by equation $B_2^0=10(\Theta_{CW}^{ab}-\Theta_{CW}^c)/3(2J-1)(2J+3)$. The calculated $B_2^0$ (= -5.8 K) closely matches with the estimating from the CEF model fitting. This further confirms the consistency of our CEF analysis.

The isothermal magnetizations \textit{M}(\textit{B}) measured along the \textit{B}$\parallel$$ab$ and \textit{B}$\parallel$$c$, as shown in Figs. \ref{Fig3}(a) and \ref{Fig3}(b), respectively. For $B$$\parallel$$ab$, the $M_{ab}$ at 2 K exhibits a rapid increase up to the critical field $B_{ab2}$, indicating the spin-flop transition. After this transition, it gradually increases and approaches a quasi-saturation value $\sim$ 0.9 $\mu_B$ for fields $B$ $>$ 2 T. On the other hand, for $B$$\parallel$$c$, the $M_c$ curves display a gradual rise without sign of saturation up to 9 T, following two field-induced metamagnetic-like transitions, as shown in Fig \ref{Fig3}(b). First transition $B_{c1}$ is more clearly evident in the field derivative of $M$($B$), shifting to lower fields with increasing temperature and disappearing completely above 10 K. This transition was not observed in previous reports of \cite{CeGaSi_2024,CeGaSi_2024_02}. The saturation magnetization values at 2 K for both directions are less from the expected value for a free Ce$^{3+}$ (g$_J$J$\mu_B$ = 2.16 $\mu_B$) ion, implying the presence of the CEF effect. As the temperature increases, magnetization value decreases for both directions. Additionally, a pronounced hysteresis is observed in the $M_{ab}$ curve, while a weaker one appears in the $M_c$ curve at 2 K [see Figs. \ref{Fig3}(c) and \ref{Fig3}(d), respectively], indicating the presence of an FM component. The magnetic hysteresis loop along \textit{B}$\parallel$$ab$ gradually diminishes with increasing temperature and vanishes around 12 K. Interestingly, the virgin $M_{ab}$ curve at 2 K shows a linear field dependence at low fields, followed by a metamagnetic-like transition around $B_{ab1} = 0.01$ T and a pronounced spin-flop transition near $B_{ab2} = 0.03$ T (see Fig. \ref{Fig3}(e)). With increasing temperature, both transitions gradually shift to lower fields and eventually disappear by 12 K, which is a typical signature of an AFM system. The presence of a well-defined hysteresis loop in $B$$\parallel$$ab$ plane and weak hysteresis along $B$$\parallel$$c$, combined with the low/high-field linear magnetization and absence of saturation, suggests the intrinsic coexistence of FM and AFM interactions. Overall magnetization behavior of CeGaSi is consistent with a canted antiferromagnets, where both FM and AFM components coexist. Similar behavior has been reported in other canted AFM systems such as as La$_2$ZnIrO$_6$ \cite{La2ZnIrO6,La2ZnIrO6_02}, Ce$_3$Cu$_3$Sb$_4$\cite{Ce3Cu3Sb4_01,Ce3Cu3Sb4_02}, BaFe$_2$Se$_4$ \cite{BaFe2Se4}, TaNi$_2$Te$_2$ \cite{TaNi2Te2}, and TbPt$_3$ \cite{TbPt3}. To precisely determine the magnetic ground state of CeGaSi, advanced microscopic techniques such as neutron diffraction, x-ray resonant magnetic scattering, or muon spin relaxation (µSR) would be essential. However, such investigations are beyond the scope of the present study.\ 

\begin{table}
	\centering 
	\caption {CEF parameters, energy levels and the corresponding wave functions of CeGaSi}
	\vskip .1cm
	\begin{tabular}{ccccc}
		\hline
		\hline \\[0.01ex]
		\multicolumn{5}{c}{CEF parameters}\\[0.5ex]
		\hline \\[0.01ex]
		$B_2^0$ (K) &~$B_4^0$ (K) &~$B_4^4$ (K)&$\lambda_i$ (mol/emu) &$\chi_0^i$ (10$^{-4}$emu/mol)\\
		-5.50 &-0.10 &5.10&$\lambda_z$ = 18.5 &$\chi_0^z$ = -5.2\\
		&     &      &$\lambda_{x,y}$ = 71.0  &$\chi_0^{x,y}$ = -1.0\\[0.1ex]
		\hline
		\hline \\[0.01ex]
	\end{tabular}
	\begin{tabular}{ccccccc}
		\multicolumn{7}{c}{Energy levels and wave functions} \\[0.5ex]
		\hline \\[0.01ex]
		~E (K)~&~$\ket{+5/2}$~&~$\ket{+3/2}$~&~$\ket{+1/2}$~&~$\ket{-1/2}$~&~$\ket{-3/2}$~&~$\ket{-5/2}$~ \\[0.5ex]
		\hline\\	
		288.1  &  -0.58     &      0     &    0     &  0   &  -0.81   &    0     \\[0.5ex]	
		288.1  &    0       &    -0.81   &    0     &  0   &    0     & -0.58    \\[0.5ex]	
		192.1  &    0       &      0     &    1     &  0   &    0     &    0    \\[0.5ex]
		192.1  &    0       &      0     &    0     &  1   &    0     &    0    \\[0.5ex]
		0      &   0.81     &      0     &    0     &  0   &  -0.58   &    0    \\[0.5ex]
		0      &    0       &    -0.58   &    0     &  0   &    0     & 0.81    \\[0.5ex]
		
		\hline
		\hline
		\label{CEF_chi}	
	\end{tabular}
\end{table}

We also calculated the magnetization data using the CEF model. The magnetization, denoted by $M_i$ ($i$ = $x$, $y$, and $z$), can be calculated by using the expression
\begin{equation}
	M_i= g_J\mu_B \sum_{n}|\bra{\Gamma_{n}} J_i \ket{\Gamma_{n}} | \frac{e^{-\beta E_n}}{Z}.  \\
	\label{MH_CEF}
\end{equation}
Here, $E_n$ signifies the eigenvalue and $\ket{\Gamma_{n}}$ represents the associated eigenfunction. These values can be obtained by diagonalizing total Hamiltonian, which is given by
\begin{equation}
	\mathcal{H} = \mathcal{H}_{CEF}-g_j \mu_BJ_i(B+\lambda_i M_i),
	\label{MH_Hem}
\end{equation}
where $\mathcal{H}_{CEF}$ is defined in Eq. \ref{CEF}, and the subsequent terms denote the Zeeman and the molecular field term, respectively. The Eq. \ref{MH_Hem} represents a nonlinear equation in which $M_i$ can be calculated through self-consistent method. The calculated magnetization data for both crystallographic directions at 2 K are displayed in Figs. \ref{Fig3}(f) and \ref{Fig3}(g) by solid violet lines, shows a deviation from experimental data. This difference might arise from the exclusion of exchange interactions in the Hamiltonian. Nevertheless, the anisotropy observed in the magnetization plots is clearly explained by the present set of crystal field parameters. Now, we have computed the saturation magnetization using Zeeman term \cite{CeIr3Ge7_CEF} 
\begin{equation}
	\label{Zeeman_term}
	\mathcal{H}_{x,z} = \textit{g}_{J}\mu_B\langle J_{x,z} \rangle\textit{B}_z,
\end{equation}
The saturation magnetizations for both crystallographic directions are determined as follows
\begin{equation}
	\label{Ms}
	\begin{split}
	&M_{B{\parallel}c} = g_J\mu_B\bra{\Gamma_{7}}J_{x,z}\ket{\Gamma_{7}}=0.96~\mu_B/Ce^{3+}.\\
&M_{B{\parallel}ab} = \textit{g}_{J}\mu_B\bra{\Gamma_{7}}\textit{J}_{x,z}\ket{\Gamma_{7}}=0.91~\mu_B/Ce^{3+}.
\end{split}
\end{equation}
Here $\Gamma$$_{7}$ (= 0.81$|$$\pm\frac{5}{2}$$\rangle$ -0.58$|$$\mp\frac{3}{2}$$\rangle$) represents the ground state wave function. The estimated saturation magnetization values are well agreement with the observed data ($M$$_{B{\parallel}c}^{obs}$ = 0.98$\mu_B$ and $M$$_{B{\parallel}ab}^{obs}$ = 0.89$\mu_B$) at 2 K.

\begin{figure}
	\includegraphics[width=6.8cm, keepaspectratio]{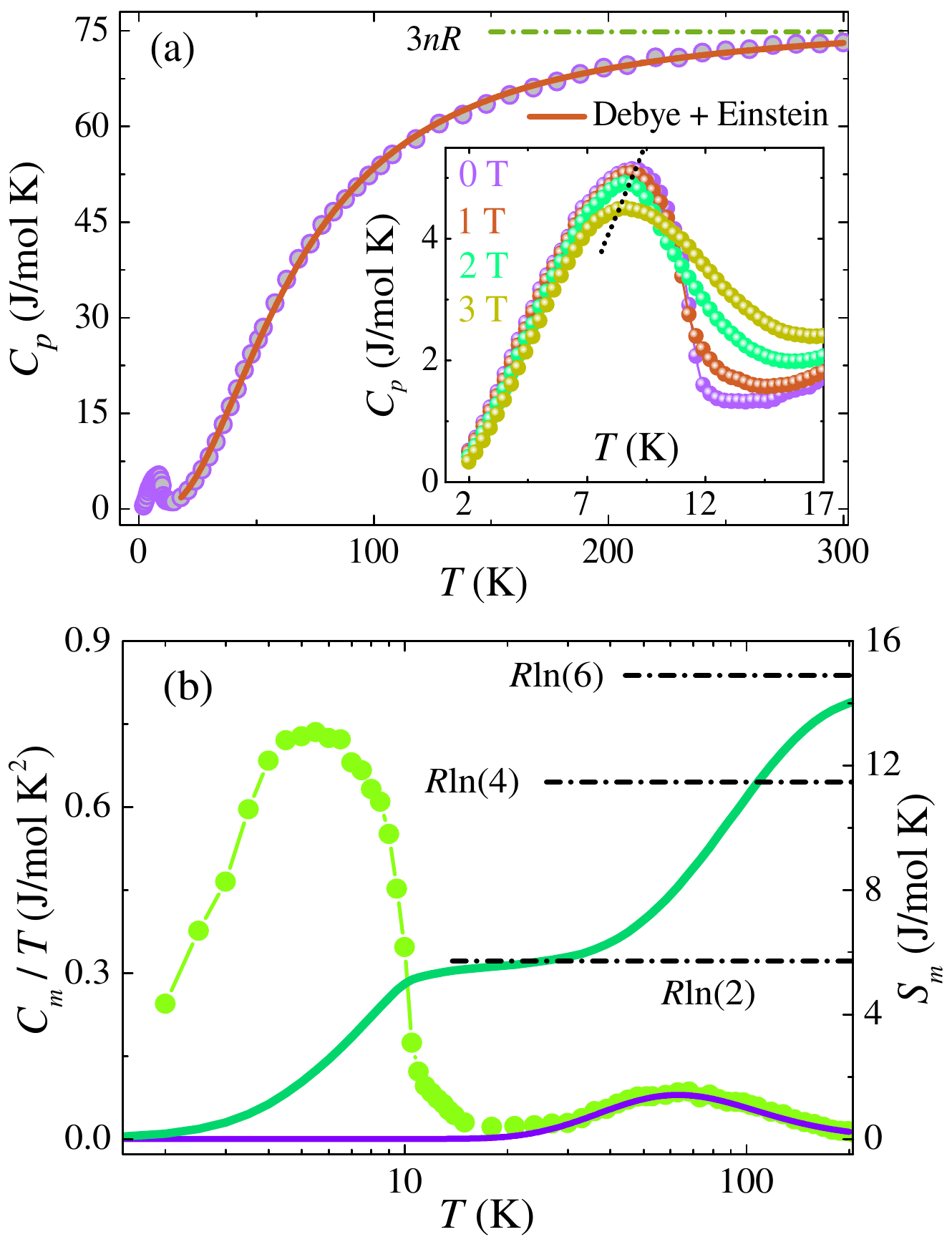}
	\caption{\label{Fig4}(a) The temperature-dependent heat capacity $C_p$($T$) of CeGaSi single crystal along with an red solid line representing the Debye-Einstein model fit. Inset shows $C_p$($T$) under various applied fields along $B$$\parallel$$c$. (b) The temperature dependence of $C_m$/$T$ with a fit of $C_m$/T data using the CEF scheme by solid violet line (right axis). The magnetic entropy $S_m$ as a function of temperature (left axis).} 
\end{figure}
\begin{figure*}
	\centering
	\includegraphics[width=17cm, keepaspectratio]{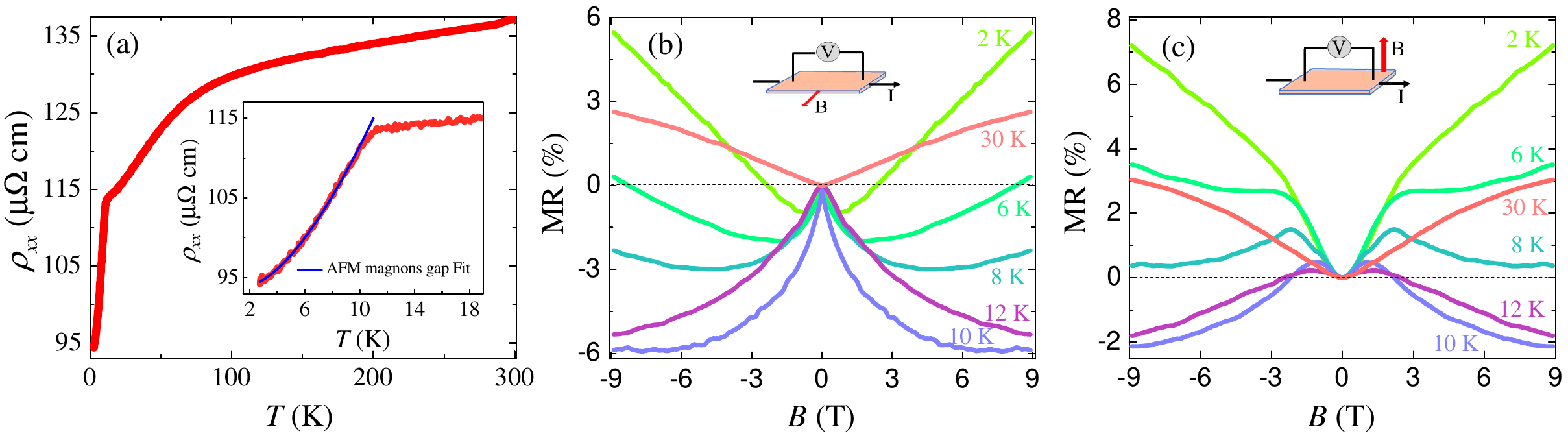}
	\caption{\label{Fig5}(a) The temperature-dependent electrical resistivity of CeGaSi single crystal in the range from 2 to 300 K. Inset displays a zoom-in view of low-temperature electrical resistivity data. The solid blue line represents a fit to the expression $\rho(T)$ = $\rho_0$ + $A$$\Delta^{3/2}$$T^{1/2}$$e^{-\Delta/T}$$\big[$1+$\frac{2}{3}$($\frac{T}{\Delta}$)+$\frac{2}{15}$($\frac{T}{\Delta}$)$^2$$\big]$. The field dependence of MRs for (b) $B$$\parallel$$ab$ and (c) $B$$\parallel$$c$, respectively. Insets of (b) and (c) show a schematic setup of MR measurements.}
\end{figure*}

\subsection{Heat capacity and entropy}
The heat capacity $\textit{C$_p$}$($\textit{T}$) of CeGaSi measured in the temperature range of 2$-$300 K at constant pressure is displayed in Fig. \ref{Fig4}(a). A $\lambda$-shaped anomaly has emerged in the $\textit{C$_p$}$($\textit{T}$) data, which corresponds to the magnetic transition at 11 K. This confirms the bulk nature of magnetic ordering in CeGaSi. In the presence of an applied magnetic field along $B$$\parallel$$c$, this peak shifts to lower temperatures and becomes progressively broader with increasing field strength, as shown in the inset of Fig. \ref{Fig4}(a). Such behavior is a signature of AFM ordering, further supporting the AFM nature of CeGaSi \cite{CeAlGe_2018,EuAg4Sb2}. Notably, the $\textit{C$_p$}$($\textit{T}$, $B$ = 0) reaches a value of $\sim$ 74 J/mol K at room temperature, which is limit in of Dulong-Petit value of $\textit{C$_p$}$ = 3$n$$\textit{R}$ = 74.84 J/mole K, where $\textit{R}$ is the universal gas constant and $n$ is the number of atoms per formula in the material. We analyzed the $\textit{C$_p$}$($\textit{T}$) data using the Debye-Einstein model in the temperature range (20$–$300 K), given follows
\begin{equation}
	C_p(\textit{T}) = \gamma \textit{T}+pC_{D}(\textit{T})+(1-p)C_{E}(\textit{T}),
	\label{Eq2}
\end{equation} 
where $p$ is the weight factor, $\gamma$$T$ is the electronic contribution, $C_D$(\textit{T}) and $C_E$(\textit{T}) represent the Debye and Einstein contributions, respectively, as denoted by the following expressions,
\begin{equation}
	C_{D}(\textit{T})=9\textit{n}R\left( \frac{\textit{T}}{\Theta_D}\right)^3\int_{0}^{\Theta_D/\textit{T}}\frac{x^4e^x}{(e^x-1)^2}dx
\end{equation}
\noindent and
\begin{equation}
	C_{E}(\textit{T})=3\textit{n}R\left( \frac{\Theta_E}{\textit{T}}\right)^2\frac{e^{\Theta_E/\textit{T}}}{(e^{\Theta_E/\textit{T}}-1)^2}.
\end{equation}
Here $\Theta_D$ and $\Theta_E$ are Debye and Einstein temperatures, respectively \cite{EuAg2Ge2}. The obtained fitting parameters are $\gamma$ = 4.58 mJ/ mol K$^2$, $\Theta_D$ = 298 K, $\Theta_E$ = 124 K, and $p$ = 0.79. The magnetic contribution to heat capacity, $C_m$($\textit{T}$), estimated by subtracting $C_p$($\textit{T}$) of the nonmagnetic reference LaGaSi from the total heat capacity of CeGaSi. The $C_m$($\textit{T}$) data in addition to magnetic transition exhibit a broad hump anomaly around 70 K, which results from the Schottky-type anomaly, as shown in Fig. \ref{Fig4}(b). To calculate a broad hump anomaly, we have used the three-level Schottky-type formula for the heat capacity, given as follows 
\begin{equation}
	\begin{split}
		C_{Sch}(T)
		=&\left(R/T^2\right)\big[\Delta_1^2 e^{-\Delta_1/T} + \Delta_2^2 e^{-\Delta_2/T}+ (\Delta_1-\Delta_2)^2\\
		& e^{-(\Delta_1 + \Delta_2)/T}\big]/(1 + e^{-\Delta_1/T} + e^{-\Delta_2/T})^{2}, 
	\end{split}
\end{equation} 
where $\Delta$$_1$ and $\Delta$$_2$ are energy gap splitting \cite{Schottky_CEF}. We have utilized the same energy level gap and degeneracy as obtained from the CEF analysis of the magnetic susceptibility data. The resultant curve reproduces the broad hump, as presented in Fig. \ref{Fig4}(b) by solid violet line, further validating of the CEF parameters and the energy level splittings. Next, the temperature dependence of magnetic entropy ($S_m$) computed by integrating $C_m$($\textit{T}$)/$\textit{T}$, as displayed on the right axis of Fig. \ref{Fig4}(b). At $T_m$, the magnetic entropy reaches a plateau with the value of 0.90$\textit{R}$ln2, suggesting a ground state of Kramers doublet. Above $T_m$, $S_m$ gradually increases and nearly reaches a saturation value of 0.94$\textit{R}$ln6 at 200 K, the reduced value of $S_m(T)$ compared with $\textit{R}$ln6 may be attributed to the inaccurate estimation of the lattice contribution to the $C_p$($\textit{T}$) \cite{EuPd2As2}.
\begin{figure*}
	\centering
	\includegraphics[width=18cm,keepaspectratio]{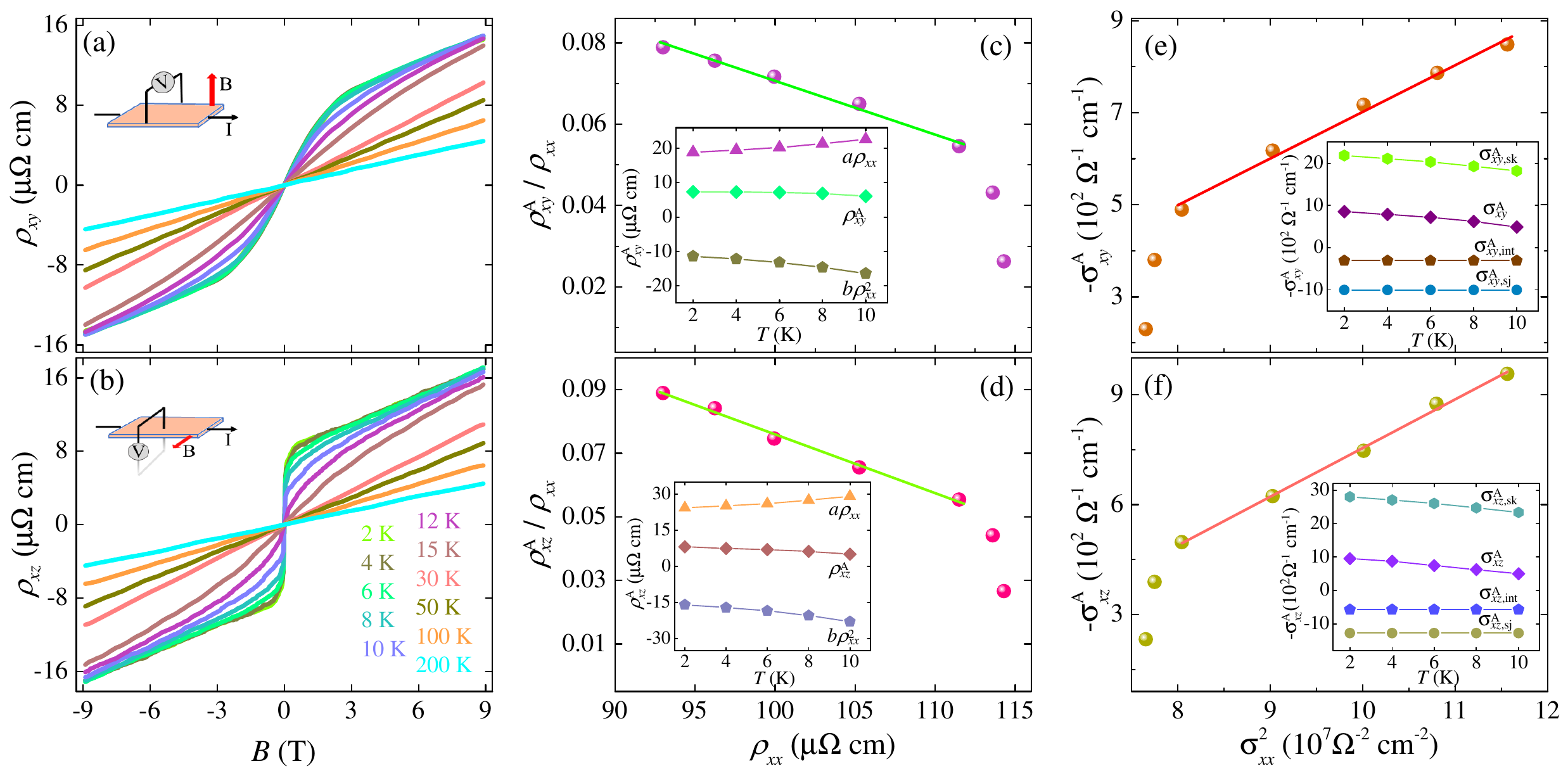}
	\caption{\label{Fig6} Field-dependent Hall resistivity (a) $\rho_{xy}$ and (b) $\rho_{xz}$ at different temperature. (c) and (d) show a linear fit to the $\rho_{xz}^A$$/$$\rho_{xx}$ and $\rho_{xy}^A$$/$$\rho_{xx}$ versus $\rho_{xx}$ curve, respectively. Insets of (c) and (d) represent different contributions in $\rho_{xy}^A$ and $\rho_{xz}^A$ with temperature. A linear fit of (e) −$\sigma_{xy}^A$ versus $\sigma_{x}^2$ and (f) −$\sigma_{xz}^A$ versus $\sigma_{xx}^2$, respectively. Insets of (e) and (f) display the different contribution (skew scattering ($\sigma_{ij, sk}^A$), intrinsic ($\sigma_{ij, int}^A$), and side jump ($\sigma_{ij, sj}^A$)) of -$\sigma_{xz}^A$ and −$\sigma_{xy}^A$ against temperature.}
\end{figure*}
\subsection{Magnetotransport}
The temperature-dependent electrical resistivity ($\rho(T)$) of CeGaSi single crystal is measured with an applied current within $ab$ plane, as shown in Fig. \ref{Fig5}(a). The $\rho$($T$) data exhibit a low residual resistivity ratio [RRR = $\rho$(300K)/$\rho$(2K)] of $\sim$ 1.45, which is likely attributed due to low carrier density and sites disorder between Ga and Si, as observed many compounds of \textit{RXY} (\textit{R} = Ce$-$Nd, \textit{X} = Al/Ga, and \textit{Y} = Si/Ge) family \cite{RGaGe_daloo,CeAlSi_2021,PrAlSi_2020,RAlGe_2019,PrGaSi}. As temperature decreases, the $\rho(T)$ decreases monotonously in a metallic manner with a broad hump around 70 K. This hump is likely a consequence of the crystal field splitting of Ce$^{3+}$ ion. With further cooling, $\rho(T)$ decreases more rapidly and shows a sharp drop around 11 K due to the freezing of spin-disorder scattering (see the inset of Fig. \ref{Fig5}(a)). A similar resistivity trend is also observed in many compounds such as CeGaGe and PrAlGe \cite{RGaGe_daloo, PrAlGe1−xSix_AHE}. Furthermore, the inset of Fig. \ref{Fig5}(a) clearly shows that $\rho(T)$ below $T_m$ is well fitted by the expression $\rho(T)$ = $\rho_0$ + $A$$\Delta^{3/2}$$T^{1/2}$$e^{-\Delta/T}$$\big[$1+$\frac{2}{3}$($\frac{T}{\Delta}$)+$\frac{2}{15}$($\frac{T}{\Delta}$)$^2$$\big]$, indicating that scattering of conduction electrons by AFM magnons dominates in this regime \cite{EuAg4Sb2}. The obtained least square fitting parameters are a residual resistivity $\rho_0$ = 93.8  $\mu$$\Omega$ cm, spin-wave gap of $\Delta$ = 8.1 K, and a coefficient $A$ = 0.26 $\mu$$\Omega$ cm K$^{-2}$. Our observed metallic behavior in $\rho(T)$ of CeGaSi contrasts with the recently studied in single crystal CeGaSi, where reported semiconductor-like behavior in $\rho(T)$ \cite{CeGaSi_2024,CeGaSi_2024_02}. This kinds of discrepancy may arise due to deviations in the stoichiometric compositions of our single crystals from previous reported, as also seen in polycrystalline CeGa$_x$Si$_{2-x}$.  However, our findings align more closely with polycrystalline CeGaSi, suggesting that the chemical composition of our sample is close to a stoichiometry ratio of CeGaSi \cite{CeSi2-xGa_1998,CeSi2-xGa_2022}.  

Figures \ref{Fig5}(b) and \ref{Fig5}(c) show MRs for two field configurations (\textit{B}$\parallel$\textit{ab} and \textit{B}$\parallel$\textit{c}) at selected temperatures. MR is defined as [$\rho$(\textit{B})$-$ $\rho$(0)]/$\rho$(0), in which $\rho$(\textit{B}) and $\rho$(0) denote the resistivity with and without magnetic field, respectively. At low temperatures, MR for \textit{B}$\parallel$\textit{ab} is negative at the low field due to the suppression of spin-disorder scattering and becomes positive at the high field due to the dominance of Lorentz force scattering. In contrast to this, MR for \textit{B}$\parallel$\textit{c} is positive in the whole measured field range. As the temperature increases, MR for both orientations become negative due to the reduction of spin-disorder scattering by the magnetic field. Maximum negative MRs for both orientations are observed near the transition temperature, where reduces the maximum spin-disorder scattering. In PM region, MRs for both orientations again becomes positive due to the dominance of Lorentz force scattering. A similar trend of MRs is also reported in several isostructural compounds like CeGaGe, PrAlSi, and PrAlGe \cite{RGaGe_daloo, PrAlSi_2020, PrAlGe_2019}
\subsection{Anomalous Hall Effect}
\label{AHE}
Next, we have performed Hall resistivity measurements ($\rho_{xz}$ and $\rho_{xy}$) of CeGaSi single crystal under an applied magnetic field along \textit{B}$\parallel$\textit{ab} and \textit{B}$\parallel$\textit{c}, as shown in Figs. \ref{Fig6}(a) and \ref{Fig6}(b), respectively. Both $\rho_{xy}$ and $\rho_{xz}$ exhibit a sharp increase at the low field, followed by a gradual linear increase at higher fields within the ordered state, which are similar to the magnetization data as shown in Figs. \ref{Fig3}(a) and \ref{Fig3}(b), respectively. This attributes to anomalous Hall response, which decreases with increasing temperature and completely disappears at $T \geq$ 20 K in both Hall resistivity data. However, anomalous Hall response in $\rho_{xy}$ is not observed in the previous report of CeGaSi \cite{CeGaSi_2024}. The Hall resistivity of magnetic materials can be expressed as $\rho_{H} =\rho_H^O (R_OB)+ \rho_H^A (R_s\mu_0M)$, where $\rho_H^O$ and $\rho_H^A$ represent the ordinary Hall effect and the AHE, respectively \cite{AHE}. 
Here $R_0$ and $R_s$ are the ordinary and anomalous Hall coefficients, respectively. The positive $R_0$ indicates dominant hole type carrier with nearly the same value in both field configurations. Using the single band model, the estimated hole density ($n_h$) = 0.5 $\times$ 10$^{21}$ cm$^{−3}$, and mobility ($\mu_{h}$) = 100 cm$^2$/V s at 30 K. As the temperature increases, the $n_h$ also increases and reaches to $\sim$2 $\times$ 10$^{21}$ cm$^{−3}$ at 300 K, while $\mu_{h}$ decreases. Similar carrier density values have been also reported in topological materials, such as PrAlGe \cite{PrAlGe_2019}, GdAgGe \cite{GdAgGe}, and YbPtBi \cite{YbPtBi}, suggesting topological nature of CeGaSi single crystals. Additionally, the value of $R_s$ is approximately an order of magnitude higher than $R_0$, indicating that the ordinary Hall effect is negligible compared to the AHE in the ordered state. Similar, $R_s$ values have been observed in several topological semimetals, such as Si-doped PrAlGe and NdAlGe \cite{NdAlGe_AHE, PrAlGe1−xSix_AHE}. 
\begin{figure*}
	\includegraphics[width=17.0cm, keepaspectratio]{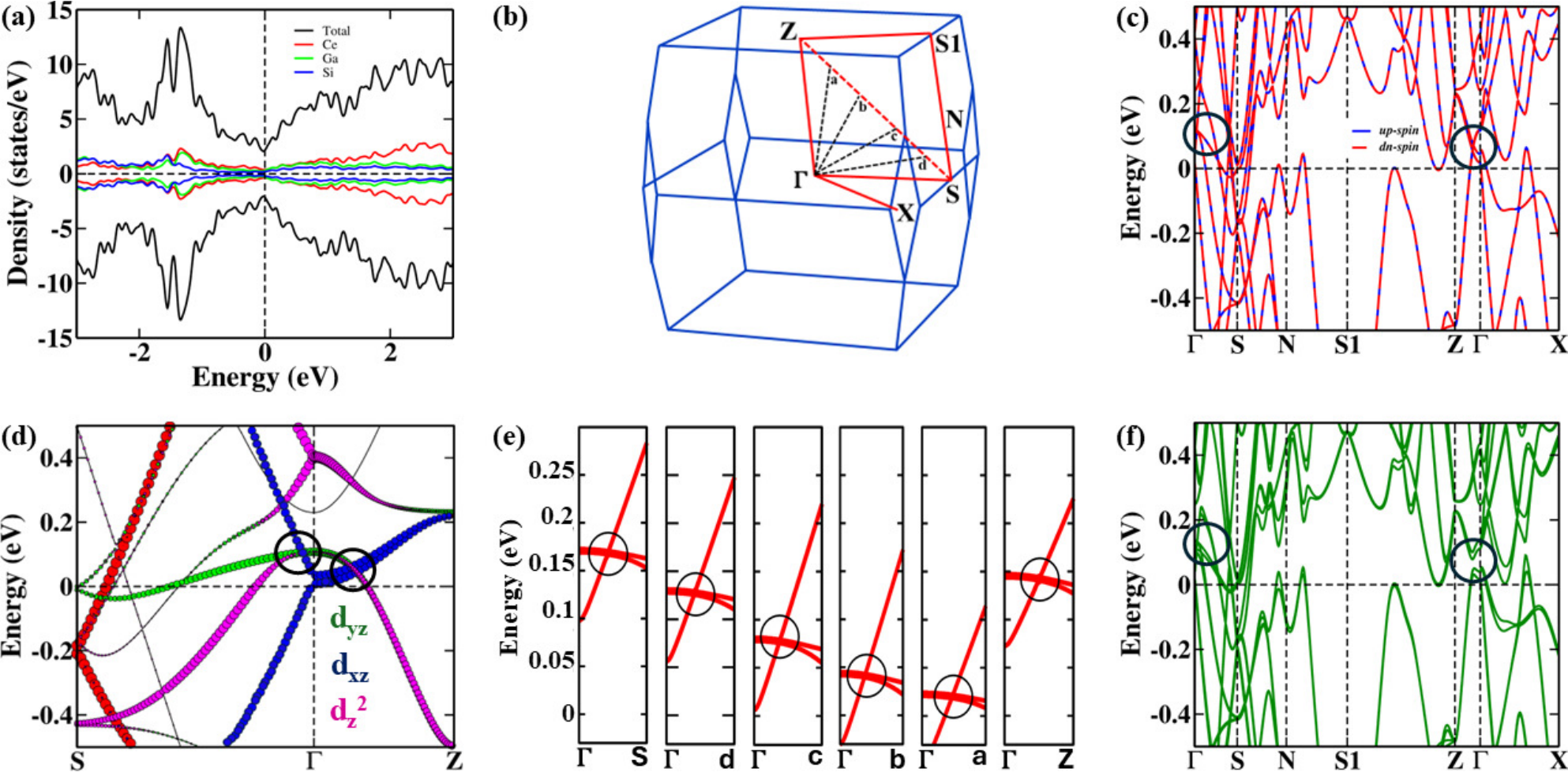}
	\caption{\label{ele}(a) Density of states plot. (b) The Brillouin zone for the $I$4$_1$\textit{md} structure, with high-symmetry points labeled. (c) Spin-resolved electronic band structure of CeGaSi, showing nodal line-like crossings (highlighted by black circles). (d) Projected bands illustrating character mixing involving Ce-\textit{d} orbitals. (e) Visualization of nodal line features along the S-G-Z path. (f) Band structure with SOC, highlighting gapped crossings (indicated by black circles).}
\end{figure*}
To unravel the underlying mechanisms driving the anomalous Hall resistivities ($\rho_{xy}^A$ and $\rho_{xz}^A$) in CeGaSi single crystal, the experimental data are fitted with the scaling laws. The anomalous Hall resistivities arise from intrinsic and extrinsic mechanisms, where the latter can arise from the skew-scattering or side-jump processes, and the former arises from non-trivial band topology. If intrinsic and side jump mechanisms contribute to $\rho_{ij}^A$, then $\rho_{ij}^A$ is proportional to the square of the longitudinal resistivity ($\rho_{ii}^2$). In contrast, for skew scattering, the $\rho_{ij}^A$ is directly proportional to $\rho_{ii}$ \cite{AHE,PrAlGe_2019,NdAlGe_AHE, PrAlSi_2020}. The scaling analysis starts with the fitting of $\rho$$_{ij}^A$/$\rho$$_{ii}$ versus $\rho$$_{ii}$ using the relation given below, as shown in Figs. \ref{Fig6}(c) and \ref{Fig6}(d).
\begin{equation} 
	\label{Hall_03}
	\rho_{ij}^A =a'\rho_{ii}+b\rho_{ii}^2,
\end{equation}
\noindent where $a'$ and $b$ coefficients parameterize the skew-scattering ($a'$$\rho$$_{ii}$) and intrinsic contribution ($b$$\rho$$_{ii}^2$), respectively. A side-jump mechanism may also contribute to $b$ through the same $\rho$$_{ii}^2$ dependence. The obtained values for these coefficients are presented inset Figs. \ref{Fig6}(c) and \ref{Fig6}(d), indicating that the skew scattering mechanism predominant in CeGaSi single crystal. This is further confirmed by scaling analysis of the anomalous Hall conductivity (AHC) as suggested in Refs. \cite{Scaling_AHE_01,Scaling_AHE_02, CeGaSi_2024}, as given follows.
\begin{equation}
	\label{Hall_04}
	-\sigma_{ij}^A =(\alpha\sigma_{ii0}^{-1}+\beta\sigma_{ii0}^{-2})\sigma_{ii}^2+m,
\end{equation}
\noindent here $\sigma_{ij}^A$ ($\sigma$$_{xy}^A$ and $\sigma$$_{xz}^A$) is the temperature-dependent AHC determined using the relation: $\sigma_{ij}^A$ ($T,B$ = 0) = −$\rho_{ij}^A/\rho_{ii}^2$ and $\sigma_{ii0}$ = 1/$\rho_{ii0}$ represents the residual conductivity. In Eq. \ref{Hall_04}, the first term contains the skew scattering along with the side jump component, whereas the second term ($m$) is the intrinsic contribution to the AHC. In our case, the fitting outcomes from Eq. \ref{Hall_04} further evident that the observed giant AHC, $\sigma_{xy}^A$ and $\sigma_{xz}^A$ (848 and 956 $\Omega^{-1}$cm$^{-1}$ at 2 K, respectively) in CeGaSi are predominantly governed by the skew scattering in both directions (see insets of Figs. \ref{Fig6}(e) and \ref{Fig6}(f)). A similarly large AHC is also observed in the isostructural compounds PrAlGe$_{1-x}$Si$_x$ ($\sim$1000-3000 $\Omega^{-1}$cm$^{-1}$), where for $x>$ 0.5, the extrinsic contribution to the AHC dominates \cite{PrAlGe1−xSix_AHE}.
\subsection{Electronic structure}
\label{ES}
CeGaSi crystallizes in a non-centrosymmetric LaPtSi-type body-centered tetragonal structure, belonging to the polar space group $I$4$_1$\textit{md}. The optimized lattice parameters are $a = b = 4.325$ \AA{} and $c = 14.443$ \AA{}, closely matching the experimental values. To determine the magnetic ground state, $1 \times 2 \times 1$ supercell was employed to explore various possible magnetic configurations, including FM and AFM arrangements. Among these, the AFM-7 with spin alignment along the [110] direction emerged as the most energetically favorable, consistent with experimental observations. The density of states for the magnetic ground state (AFM-7) is presented in Fig. \ref{ele}(a), showcasing the overall contributions of the constituent atoms in CeGaSi, as well as their individual contributions. The metallic nature of the compound is evident from the overlap between the valence and conduction bands at the Fermi level (\( E_F \)). Notably, the valence band is primarily formed by an almost equal contribution from the Ce, Ga, and Si atoms. On the other hand, the conduction band is predominantly composed of contributions from Ce atoms, with smaller but comparable contributions from Ga and Si atoms. This distinct distribution emphasizes the importance of specific atoms in determining the electronic characteristics of the compound, underlining their role in shaping the material’s electronic structure. The Brillouin zone for the $I$4$_1$\textit{md} structure, along with the high-symmetry points and the path along which the electronic band structure was calculated, is shown in Fig. \ref{ele}(b). The spin-polarized band structure, as depicted in Fig. \ref{ele}(c), reveals the presence of nodal line-like crossings between the $\Gamma$-S path at an energy of 0.1 eV and the Z-$\Gamma$ path at an energy of 0.06 eV, marked by black circles. To explore the topological aspects of the crossings, the projected bands along the path S-$\Gamma$-Z is plotted as depicted in Fig. \ref{ele}(d). 

An interesting feature of the band structure is the character mixing occurring between the \(d\)-orbitals of Ce atoms, particularly the \(d_{yz}\), \(d_{xz}\), and $d_{z^2}$ orbitals. This mixing leads to  band inversion, which is indicative of the material's non-trivial topological properties. The band hybridization, which involves the Ce-($d_{yz}$, $d_{xz}$, and $d_{z^2}$) states, is responsible for the formation of the nodal line-like dispersion. To further investigate this, several band pathways were selected between S (0.0, 0.5, 0.0) and Z (0.0, 0.0, 0.5), with $\Gamma$ (0.0, 0.0, 0.0) as the center as shown in Fig. \ref{ele}(b), lying in the \(k_x = 0\) plane. The dispersion along all the band path ways reveal similar crossings, as illustrated in Fig. \ref{ele}(e).

When spin-orbit coupling (SOC) is introduced along the [110] direction, the degeneracy at the crossing points are lifted, resulting in gaps in the electronic band dispersion, as depicted in Fig. \ref{ele}(f). This demonstrates the significant role of SOC in shaping the topological properties of the material. The inclusion of SOC disrupts several symmetries of the system. First, SOC breaks the fourfold rotational symmetry (\(C_4\)) around the \(c\) axis because the spin-orbit interaction along a specific direction, such as [110], does not commute with \(C_4\) rotations. This reduces the overall rotational symmetry. Furthermore, mirror symmetries perpendicular to the [110] direction are also violated, as the spin-orbit interaction modifies the symmetry of the electronic states. Although IS is inherently absent due to the non-centrosymmetric nature of the crystal structure, SOC amplifies this asymmetry by introducing spin-dependent effects that are not inversion symmetric. The most pronounced impact of SOC is the opening of the gaps at the  crossings, particularly along the $\Gamma$-S and Z-$\Gamma$ paths, as shown in Fig. \ref{ele}(f). This gap arises due to the breaking of mirror symmetry perpendicular to the [110] direction and signals the emergence of non-trivial topological characteristics in the material.
\section{Summary}
\label{V}
In summary, we systematically investigated the magnetic, thermodynamic, and magnetotransport properties of CeGaSi single crystal, which crystallizes in tetragonal LaPtSi-type structure. CeGaSi orders magnetically around $T_m$ $\sim$ 11 K, which was confirmed through the magnetic susceptibility, heat capacity, and electrical resistivity data. Our CEF analysis on magnetic susceptibility, magnetization, and heat capacity data of CeGaSi shows the six degenerate ground states of the Ce$^{3+}$ ion split into three doublets. The $M$($B$) data exhibit multiple metamagnetic-like transitions at low temperatures, which progressively weaken and shift to lower fields with increasing temperature, eventually disappearing above 10 K. The field-dependent MRs for both directions are positive at low temperatures and high fields. MRs switch to negative near $T_m$ and again become positive in the PM region. We observed a large AHE in both field directions, predominantly attributed to the contribution of skew scattering. Furthermore, first-principles calculations were performed to examine the electronic and topological properties of CeGaSi within its magnetic ground state. The analysis uncovered nodal line-like features along the S-G-Z path lying in the \(k_x = 0\) plane, thereby classifying CeGaSi as a nodal-line metal.


\section{Acknowledgments}
We acknowledge IIT Kanpur and the Department of Science and Technology, India, [Order No.
DST/NM/TUE/QM-06/2019 (G)] for financial support. D. R. thanks the Indo-French Centre for the Promotion of Advanced Research (IFCPAR) for a postdoctoral fellowship (Project No. 7108-2). A. CV and V. K. express their gratitude for the support and resources extended by IIT Hyderabad and the National Supercomputing Mission through the 'PARAM SEVA' by CDAC, for computational resources. V.K. appreciates the financial support received from DST-FIST (SR/FST/PSI-215/2016).  

\bibliography{CeGaSi_ref}

\begin{thebibliography}{56}%
\makeatletter
\providecommand \@ifxundefined [1]{%
 \@ifx{#1\undefined}
}%
\providecommand \@ifnum [1]{%
 \ifnum #1\expandafter \@firstoftwo
 \else \expandafter \@secondoftwo
 \fi
}%
\providecommand \@ifx [1]{%
 \ifx #1\expandafter \@firstoftwo
 \else \expandafter \@secondoftwo
 \fi
}%
\providecommand \natexlab [1]{#1}%
\providecommand \enquote  [1]{``#1''}%
\providecommand \bibnamefont  [1]{#1}%
\providecommand \bibfnamefont [1]{#1}%
\providecommand \citenamefont [1]{#1}%
\providecommand \href@noop [0]{\@secondoftwo}%
\providecommand \href [0]{\begingroup \@sanitize@url \@href}%
\providecommand \@href[1]{\@@startlink{#1}\@@href}%
\providecommand \@@href[1]{\endgroup#1\@@endlink}%
\providecommand \@sanitize@url [0]{\catcode `\\12\catcode `\$12\catcode
  `\&12\catcode `\#12\catcode `\^12\catcode `\_12\catcode `\%12\relax}%
\providecommand \@@startlink[1]{}%
\providecommand \@@endlink[0]{}%
\providecommand \url  [0]{\begingroup\@sanitize@url \@url }%
\providecommand \@url [1]{\endgroup\@href {#1}{\urlprefix }}%
\providecommand \urlprefix  [0]{URL }%
\providecommand \Eprint [0]{\href }%
\providecommand \doibase [0]{https://doi.org/}%
\providecommand \selectlanguage [0]{\@gobble}%
\providecommand \bibinfo  [0]{\@secondoftwo}%
\providecommand \bibfield  [0]{\@secondoftwo}%
\providecommand \translation [1]{[#1]}%
\providecommand \BibitemOpen [0]{}%
\providecommand \bibitemStop [0]{}%
\providecommand \bibitemNoStop [0]{.\EOS\space}%
\providecommand \EOS [0]{\spacefactor3000\relax}%
\providecommand \BibitemShut  [1]{\csname bibitem#1\endcsname}%
\let\auto@bib@innerbib\@empty
\bibitem [{\citenamefont {Kim}\ \emph {et~al.}(2003)\citenamefont {Kim},
  \citenamefont {Echizen}, \citenamefont {Umeo}, \citenamefont {Kobayashi},
  \citenamefont {Sera}, \citenamefont {Salamakha}, \citenamefont {Sologub},
  \citenamefont {Takabatake}, \citenamefont {Chen}, \citenamefont {Tayama},
  \citenamefont {Sakakibara}, \citenamefont {Jung},\ and\ \citenamefont
  {Maple}}]{CeRhSn}%
  \BibitemOpen
  \bibfield  {author} {\bibinfo {author} {\bibfnamefont {M.~S.}\ \bibnamefont
  {Kim}}, \bibinfo {author} {\bibfnamefont {Y.}~\bibnamefont {Echizen}},
  \bibinfo {author} {\bibfnamefont {K.}~\bibnamefont {Umeo}}, \bibinfo {author}
  {\bibfnamefont {S.}~\bibnamefont {Kobayashi}}, \bibinfo {author}
  {\bibfnamefont {M.}~\bibnamefont {Sera}}, \bibinfo {author} {\bibfnamefont
  {P.~S.}\ \bibnamefont {Salamakha}}, \bibinfo {author} {\bibfnamefont {O.~L.}\
  \bibnamefont {Sologub}}, \bibinfo {author} {\bibfnamefont {T.}~\bibnamefont
  {Takabatake}}, \bibinfo {author} {\bibfnamefont {X.}~\bibnamefont {Chen}},
  \bibinfo {author} {\bibfnamefont {T.}~\bibnamefont {Tayama}}, \bibinfo
  {author} {\bibfnamefont {T.}~\bibnamefont {Sakakibara}}, \bibinfo {author}
  {\bibfnamefont {M.~H.}\ \bibnamefont {Jung}},\ and\ \bibinfo {author}
  {\bibfnamefont {M.~B.}\ \bibnamefont {Maple}},\ }\bibfield  {title} {\bibinfo
  {title} {Low-temperature anomalies in magnetic, transport, and thermal
  properties of single-crystal {CeRhSn} with valence fluctuations},\ }\href
  {https://doi.org/10.1103/PhysRevB.68.054416} {\bibfield  {journal} {\bibinfo
  {journal} {Phys. Rev. B}\ }\textbf {\bibinfo {volume} {68}},\ \bibinfo
  {pages} {054416} (\bibinfo {year} {2003})}\BibitemShut {NoStop}%
\bibitem [{\citenamefont {Ram}\ \emph {et~al.}(2023{\natexlab{a}})\citenamefont
  {Ram}, \citenamefont {Malick}, \citenamefont {Hossain},\ and\ \citenamefont
  {Kaczorowski}}]{RGaGe_daloo}%
  \BibitemOpen
  \bibfield  {author} {\bibinfo {author} {\bibfnamefont {D.}~\bibnamefont
  {Ram}}, \bibinfo {author} {\bibfnamefont {S.}~\bibnamefont {Malick}},
  \bibinfo {author} {\bibfnamefont {Z.}~\bibnamefont {Hossain}},\ and\ \bibinfo
  {author} {\bibfnamefont {D.}~\bibnamefont {Kaczorowski}},\ }\bibfield
  {title} {\bibinfo {title} {Magnetic, thermodynamic, and magnetotransport
  properties of {CeGaGe} and {PrGaGe} single crystals},\ }\href
  {https://doi.org/10.1103/PhysRevB.108.024428} {\bibfield  {journal} {\bibinfo
   {journal} {Phys. Rev. B}\ }\textbf {\bibinfo {volume} {108}},\ \bibinfo
  {pages} {024428} (\bibinfo {year} {2023}{\natexlab{a}})}\BibitemShut
  {NoStop}%
\bibitem [{\citenamefont {Ram}\ \emph {et~al.}(2024)\citenamefont {Ram},
  \citenamefont {Joshi},\ and\ \citenamefont {Hossain}}]{NdGaGe_daloo}%
  \BibitemOpen
  \bibfield  {author} {\bibinfo {author} {\bibfnamefont {D.}~\bibnamefont
  {Ram}}, \bibinfo {author} {\bibfnamefont {L.}~\bibnamefont {Joshi}},\ and\
  \bibinfo {author} {\bibfnamefont {Z.}~\bibnamefont {Hossain}},\ }\bibfield
  {title} {\bibinfo {title} {{Crystalline electric field and large anomalous
  Hall effect in NdGaGe single crystals}},\ }\href
  {https://doi.org/https://doi.org/10.1016/j.jmmm.2024.172326} {\bibfield
  {journal} {\bibinfo  {journal} {Journal of Magnetism and Magnetic Materials}\
  }\textbf {\bibinfo {volume} {605}},\ \bibinfo {pages} {172326} (\bibinfo
  {year} {2024})}\BibitemShut {NoStop}%
\bibitem [{\citenamefont {Kang}\ \emph {et~al.}(1998)\citenamefont {Kang},
  \citenamefont {Olson}, \citenamefont {Inada}, \citenamefont
  {\ifmmode~\bar{O}\else \={O}\fi{}nuki}, \citenamefont {Kwon},\ and\
  \citenamefont {Min}}]{CeNiSn}%
  \BibitemOpen
  \bibfield  {author} {\bibinfo {author} {\bibfnamefont {J.-S.}\ \bibnamefont
  {Kang}}, \bibinfo {author} {\bibfnamefont {C.~G.}\ \bibnamefont {Olson}},
  \bibinfo {author} {\bibfnamefont {Y.}~\bibnamefont {Inada}}, \bibinfo
  {author} {\bibfnamefont {Y.}~\bibnamefont {\ifmmode~\bar{O}\else
  \={O}\fi{}nuki}}, \bibinfo {author} {\bibfnamefont {S.~K.}\ \bibnamefont
  {Kwon}},\ and\ \bibinfo {author} {\bibfnamefont {B.~I.}\ \bibnamefont
  {Min}},\ }\bibfield  {title} {\bibinfo {title} {Valence-band photoemission
  study of single crystalline {CeNiSn}},\ }\href
  {https://doi.org/10.1103/PhysRevB.58.4426} {\bibfield  {journal} {\bibinfo
  {journal} {Phys. Rev. B}\ }\textbf {\bibinfo {volume} {58}},\ \bibinfo
  {pages} {4426} (\bibinfo {year} {1998})}\BibitemShut {NoStop}%
\bibitem [{\citenamefont {Nam}\ \emph {et~al.}(2019)\citenamefont {Nam},
  \citenamefont {Kang}, \citenamefont {Ryu}, \citenamefont {Kim}, \citenamefont
  {Kim}, \citenamefont {Kim},\ and\ \citenamefont {Min}}]{CeNiSn_theory}%
  \BibitemOpen
  \bibfield  {author} {\bibinfo {author} {\bibfnamefont {T.-S.}\ \bibnamefont
  {Nam}}, \bibinfo {author} {\bibfnamefont {C.-J.}\ \bibnamefont {Kang}},
  \bibinfo {author} {\bibfnamefont {D.-C.}\ \bibnamefont {Ryu}}, \bibinfo
  {author} {\bibfnamefont {J.}~\bibnamefont {Kim}}, \bibinfo {author}
  {\bibfnamefont {H.}~\bibnamefont {Kim}}, \bibinfo {author} {\bibfnamefont
  {K.}~\bibnamefont {Kim}},\ and\ \bibinfo {author} {\bibfnamefont {B.~I.}\
  \bibnamefont {Min}},\ }\bibfield  {title} {\bibinfo {title} {Topological bulk
  band structures of the hourglass and {Dirac} nodal-loop types in {Ce} {Kondo}
  systems: {CeNiSn}, {CeRhAs}, and {CeRhSb}},\ }\href
  {https://doi.org/10.1103/PhysRevB.99.125115} {\bibfield  {journal} {\bibinfo
  {journal} {Phys. Rev. B}\ }\textbf {\bibinfo {volume} {99}},\ \bibinfo
  {pages} {125115} (\bibinfo {year} {2019})}\BibitemShut {NoStop}%
\bibitem [{\citenamefont {Alam}\ \emph {et~al.}(2023)\citenamefont {Alam},
  \citenamefont {Fakhredine}, \citenamefont {Ahmad}, \citenamefont {Tanwar},
  \citenamefont {Yang}, \citenamefont {Tafti}, \citenamefont {Cuono},
  \citenamefont {Islam}, \citenamefont {Singh}, \citenamefont {Lynnyk},
  \citenamefont {Autieri},\ and\ \citenamefont {Matusiak}}]{CeAlSi_Hall02}%
  \BibitemOpen
  \bibfield  {author} {\bibinfo {author} {\bibfnamefont {M.~S.}\ \bibnamefont
  {Alam}}, \bibinfo {author} {\bibfnamefont {A.}~\bibnamefont {Fakhredine}},
  \bibinfo {author} {\bibfnamefont {M.}~\bibnamefont {Ahmad}}, \bibinfo
  {author} {\bibfnamefont {P.~K.}\ \bibnamefont {Tanwar}}, \bibinfo {author}
  {\bibfnamefont {H.-Y.}\ \bibnamefont {Yang}}, \bibinfo {author}
  {\bibfnamefont {F.}~\bibnamefont {Tafti}}, \bibinfo {author} {\bibfnamefont
  {G.}~\bibnamefont {Cuono}}, \bibinfo {author} {\bibfnamefont
  {R.}~\bibnamefont {Islam}}, \bibinfo {author} {\bibfnamefont
  {B.}~\bibnamefont {Singh}}, \bibinfo {author} {\bibfnamefont
  {A.}~\bibnamefont {Lynnyk}}, \bibinfo {author} {\bibfnamefont
  {C.}~\bibnamefont {Autieri}},\ and\ \bibinfo {author} {\bibfnamefont
  {M.}~\bibnamefont {Matusiak}},\ }\bibfield  {title} {\bibinfo {title} {{Sign
  change of anomalous Hall effect and anomalous Nernst effect in the Weyl
  semimetal CeAlSi}},\ }\href {https://doi.org/10.1103/PhysRevB.107.085102}
  {\bibfield  {journal} {\bibinfo  {journal} {Phys. Rev. B}\ }\textbf {\bibinfo
  {volume} {107}},\ \bibinfo {pages} {085102} (\bibinfo {year}
  {2023})}\BibitemShut {NoStop}%
\bibitem [{\citenamefont {Piva}\ \emph
  {et~al.}(2023{\natexlab{a}})\citenamefont {Piva}, \citenamefont {Souza},
  \citenamefont {Brousseau-Couture}, \citenamefont {Sorn}, \citenamefont
  {Pakuszewski}, \citenamefont {John}, \citenamefont {Adriano}, \citenamefont
  {C\^ot\'e}, \citenamefont {Pagliuso}, \citenamefont {Paramekanti},\ and\
  \citenamefont {Nicklas}}]{CeAlSi_loop_Hall_effect}%
  \BibitemOpen
  \bibfield  {author} {\bibinfo {author} {\bibfnamefont {M.~M.}\ \bibnamefont
  {Piva}}, \bibinfo {author} {\bibfnamefont {J.~C.}\ \bibnamefont {Souza}},
  \bibinfo {author} {\bibfnamefont {V.}~\bibnamefont {Brousseau-Couture}},
  \bibinfo {author} {\bibfnamefont {S.}~\bibnamefont {Sorn}}, \bibinfo {author}
  {\bibfnamefont {K.~R.}\ \bibnamefont {Pakuszewski}}, \bibinfo {author}
  {\bibfnamefont {J.~K.}\ \bibnamefont {John}}, \bibinfo {author}
  {\bibfnamefont {C.}~\bibnamefont {Adriano}}, \bibinfo {author} {\bibfnamefont
  {M.}~\bibnamefont {C\^ot\'e}}, \bibinfo {author} {\bibfnamefont {P.~G.}\
  \bibnamefont {Pagliuso}}, \bibinfo {author} {\bibfnamefont {A.}~\bibnamefont
  {Paramekanti}},\ and\ \bibinfo {author} {\bibfnamefont {M.}~\bibnamefont
  {Nicklas}},\ }\bibfield  {title} {\bibinfo {title} {{Topological features in
  the ferromagnetic Weyl semimetal CeAlSi: Role of domain walls}},\ }\href
  {https://doi.org/10.1103/PhysRevResearch.5.013068} {\bibfield  {journal}
  {\bibinfo  {journal} {Phys. Rev. Res.}\ }\textbf {\bibinfo {volume} {5}},\
  \bibinfo {pages} {013068} (\bibinfo {year} {2023}{\natexlab{a}})}\BibitemShut
  {NoStop}%
\bibitem [{\citenamefont {Piva}\ \emph
  {et~al.}(2023{\natexlab{b}})\citenamefont {Piva}, \citenamefont {Souza},
  \citenamefont {Lombardi}, \citenamefont {Pakuszewski}, \citenamefont
  {Adriano}, \citenamefont {Pagliuso},\ and\ \citenamefont
  {Nicklas}}]{CeAlGe_Hall}%
  \BibitemOpen
  \bibfield  {author} {\bibinfo {author} {\bibfnamefont {M.~M.}\ \bibnamefont
  {Piva}}, \bibinfo {author} {\bibfnamefont {J.~C.}\ \bibnamefont {Souza}},
  \bibinfo {author} {\bibfnamefont {G.~A.}\ \bibnamefont {Lombardi}}, \bibinfo
  {author} {\bibfnamefont {K.~R.}\ \bibnamefont {Pakuszewski}}, \bibinfo
  {author} {\bibfnamefont {C.}~\bibnamefont {Adriano}}, \bibinfo {author}
  {\bibfnamefont {P.~G.}\ \bibnamefont {Pagliuso}},\ and\ \bibinfo {author}
  {\bibfnamefont {M.}~\bibnamefont {Nicklas}},\ }\bibfield  {title} {\bibinfo
  {title} {{Topological Hall effect in CeAlGe}},\ }\href
  {https://doi.org/10.1103/PhysRevMaterials.7.074204} {\bibfield  {journal}
  {\bibinfo  {journal} {Phys. Rev. Mater.}\ }\textbf {\bibinfo {volume} {7}},\
  \bibinfo {pages} {074204} (\bibinfo {year} {2023}{\natexlab{b}})}\BibitemShut
  {NoStop}%
\bibitem [{\citenamefont {Puphal}\ \emph {et~al.}(2020)\citenamefont {Puphal},
  \citenamefont {Pomjakushin}, \citenamefont {Kanazawa}, \citenamefont
  {Ukleev}, \citenamefont {Gawryluk}, \citenamefont {Ma}, \citenamefont
  {Naamneh}, \citenamefont {Plumb}, \citenamefont {Keller}, \citenamefont
  {Cubitt}, \citenamefont {Pomjakushina},\ and\ \citenamefont
  {White}}]{CeAlGe_THE}%
  \BibitemOpen
  \bibfield  {author} {\bibinfo {author} {\bibfnamefont {P.}~\bibnamefont
  {Puphal}}, \bibinfo {author} {\bibfnamefont {V.}~\bibnamefont {Pomjakushin}},
  \bibinfo {author} {\bibfnamefont {N.}~\bibnamefont {Kanazawa}}, \bibinfo
  {author} {\bibfnamefont {V.}~\bibnamefont {Ukleev}}, \bibinfo {author}
  {\bibfnamefont {D.~J.}\ \bibnamefont {Gawryluk}}, \bibinfo {author}
  {\bibfnamefont {J.}~\bibnamefont {Ma}}, \bibinfo {author} {\bibfnamefont
  {M.}~\bibnamefont {Naamneh}}, \bibinfo {author} {\bibfnamefont {N.~C.}\
  \bibnamefont {Plumb}}, \bibinfo {author} {\bibfnamefont {L.}~\bibnamefont
  {Keller}}, \bibinfo {author} {\bibfnamefont {R.}~\bibnamefont {Cubitt}},
  \bibinfo {author} {\bibfnamefont {E.}~\bibnamefont {Pomjakushina}},\ and\
  \bibinfo {author} {\bibfnamefont {J.~S.}\ \bibnamefont {White}},\ }\bibfield
  {title} {\bibinfo {title} {{Topological Magnetic Phase in the Candidate Weyl
  Semimetal CeAlGe}},\ }\href {https://doi.org/10.1103/PhysRevLett.124.017202}
  {\bibfield  {journal} {\bibinfo  {journal} {Phys. Rev. Lett.}\ }\textbf
  {\bibinfo {volume} {124}},\ \bibinfo {pages} {017202} (\bibinfo {year}
  {2020})}\BibitemShut {NoStop}%
\bibitem [{\citenamefont {Yang}\ \emph {et~al.}(2021)\citenamefont {Yang},
  \citenamefont {Singh}, \citenamefont {Gaudet}, \citenamefont {Lu},
  \citenamefont {Huang}, \citenamefont {Chiu}, \citenamefont {Huang},
  \citenamefont {Wang}, \citenamefont {Bahrami}, \citenamefont {Xu},
  \citenamefont {Franklin}, \citenamefont {Sochnikov}, \citenamefont {Graf},
  \citenamefont {Xu}, \citenamefont {Zhao}, \citenamefont {Hoffman},
  \citenamefont {Lin}, \citenamefont {Torchinsky}, \citenamefont {Broholm},
  \citenamefont {Bansil},\ and\ \citenamefont {Tafti}}]{CeAlSi_2021}%
  \BibitemOpen
  \bibfield  {author} {\bibinfo {author} {\bibfnamefont {H.-Y.}\ \bibnamefont
  {Yang}}, \bibinfo {author} {\bibfnamefont {B.}~\bibnamefont {Singh}},
  \bibinfo {author} {\bibfnamefont {J.}~\bibnamefont {Gaudet}}, \bibinfo
  {author} {\bibfnamefont {B.}~\bibnamefont {Lu}}, \bibinfo {author}
  {\bibfnamefont {C.-Y.}\ \bibnamefont {Huang}}, \bibinfo {author}
  {\bibfnamefont {W.-C.}\ \bibnamefont {Chiu}}, \bibinfo {author}
  {\bibfnamefont {S.-M.}\ \bibnamefont {Huang}}, \bibinfo {author}
  {\bibfnamefont {B.}~\bibnamefont {Wang}}, \bibinfo {author} {\bibfnamefont
  {F.}~\bibnamefont {Bahrami}}, \bibinfo {author} {\bibfnamefont
  {B.}~\bibnamefont {Xu}}, \bibinfo {author} {\bibfnamefont {J.}~\bibnamefont
  {Franklin}}, \bibinfo {author} {\bibfnamefont {I.}~\bibnamefont {Sochnikov}},
  \bibinfo {author} {\bibfnamefont {D.~E.}\ \bibnamefont {Graf}}, \bibinfo
  {author} {\bibfnamefont {G.}~\bibnamefont {Xu}}, \bibinfo {author}
  {\bibfnamefont {Y.}~\bibnamefont {Zhao}}, \bibinfo {author} {\bibfnamefont
  {C.~M.}\ \bibnamefont {Hoffman}}, \bibinfo {author} {\bibfnamefont
  {H.}~\bibnamefont {Lin}}, \bibinfo {author} {\bibfnamefont {D.~H.}\
  \bibnamefont {Torchinsky}}, \bibinfo {author} {\bibfnamefont {C.~L.}\
  \bibnamefont {Broholm}}, \bibinfo {author} {\bibfnamefont {A.}~\bibnamefont
  {Bansil}},\ and\ \bibinfo {author} {\bibfnamefont {F.}~\bibnamefont
  {Tafti}},\ }\bibfield  {title} {\bibinfo {title} {{Noncollinear ferromagnetic
  Weyl semimetal with anisotropic anomalous Hall effect}},\ }\href
  {https://doi.org/10.1103/PhysRevB.103.115143} {\bibfield  {journal} {\bibinfo
   {journal} {Phys. Rev. B}\ }\textbf {\bibinfo {volume} {103}},\ \bibinfo
  {pages} {115143} (\bibinfo {year} {2021})}\BibitemShut {NoStop}%
\bibitem [{\citenamefont {Lu}\ \emph {et~al.}()\citenamefont {Lu},
  \citenamefont {Ji}, \citenamefont {Wang}, \citenamefont {Yan}, \citenamefont
  {Chen}, \citenamefont {Zhao}, \citenamefont {Zang},\ and\ \citenamefont
  {Liu}}]{CeAlSi_2023}%
  \BibitemOpen
  \bibfield  {author} {\bibinfo {author} {\bibfnamefont {H.}~\bibnamefont
  {Lu}}, \bibinfo {author} {\bibfnamefont {W.}~\bibnamefont {Ji}}, \bibinfo
  {author} {\bibfnamefont {X.}~\bibnamefont {Wang}}, \bibinfo {author}
  {\bibfnamefont {E.}~\bibnamefont {Yan}}, \bibinfo {author} {\bibfnamefont
  {G.}~\bibnamefont {Chen}}, \bibinfo {author} {\bibfnamefont {Y.}~\bibnamefont
  {Zhao}}, \bibinfo {author} {\bibfnamefont {H.}~\bibnamefont {Zang}},\ and\
  \bibinfo {author} {\bibfnamefont {W.}~\bibnamefont {Liu}},\ }\bibfield
  {title} {\bibinfo {title} {{Significant Anomalous Hall Coefficients Observed
  in Magnetic Weyl Semimetal and Prospective Candidate RAlSi}},\ }\href
  {http://dx.doi.org/10.2139/ssrn.4508621} {\bibinfo  {journal} {Available at
  SSRN 4508621}\ }\BibitemShut {NoStop}%
\bibitem [{\citenamefont {Gong}\ \emph {et~al.}(2024)\citenamefont {Gong},
  \citenamefont {Wang}, \citenamefont {Han}, \citenamefont {Zeng},
  \citenamefont {Ma}, \citenamefont {Wang}, \citenamefont {Lin}, \citenamefont
  {Wang},\ and\ \citenamefont {Xia}}]{CeGaSi_2024}%
  \BibitemOpen
\bibfield  {journal} {  }\bibfield  {author} {\bibinfo {author} {\bibfnamefont
  {J.}~\bibnamefont {Gong}}, \bibinfo {author} {\bibfnamefont {H.}~\bibnamefont
  {Wang}}, \bibinfo {author} {\bibfnamefont {K.}~\bibnamefont {Han}}, \bibinfo
  {author} {\bibfnamefont {X.-Y.}\ \bibnamefont {Zeng}}, \bibinfo {author}
  {\bibfnamefont {X.-P.}\ \bibnamefont {Ma}}, \bibinfo {author} {\bibfnamefont
  {Y.-T.}\ \bibnamefont {Wang}}, \bibinfo {author} {\bibfnamefont {J.-F.}\
  \bibnamefont {Lin}}, \bibinfo {author} {\bibfnamefont {X.-Y.}\ \bibnamefont
  {Wang}},\ and\ \bibinfo {author} {\bibfnamefont {T.-L.}\ \bibnamefont
  {Xia}},\ }\bibfield  {title} {\bibinfo {title} {{Anomalous Hall effect in an
  antiferromagnetic CeGaSi single crystal}},\ }\href
  {https://doi.org/10.1103/PhysRevB.109.024434} {\bibfield  {journal} {\bibinfo
   {journal} {Phys. Rev. B}\ }\textbf {\bibinfo {volume} {109}},\ \bibinfo
  {pages} {024434} (\bibinfo {year} {2024})}\BibitemShut {NoStop}%
\bibitem [{\citenamefont {Kabeya}\ \emph {et~al.}(2022)\citenamefont {Kabeya},
  \citenamefont {Takahara}, \citenamefont {Arisumi}, \citenamefont {Kimura},
  \citenamefont {Araki}, \citenamefont {Katoh},\ and\ \citenamefont
  {Ochiai}}]{Ce2Pd2Pb_CEF}%
  \BibitemOpen
  \bibfield  {author} {\bibinfo {author} {\bibfnamefont {N.}~\bibnamefont
  {Kabeya}}, \bibinfo {author} {\bibfnamefont {S.}~\bibnamefont {Takahara}},
  \bibinfo {author} {\bibfnamefont {T.}~\bibnamefont {Arisumi}}, \bibinfo
  {author} {\bibfnamefont {S.}~\bibnamefont {Kimura}}, \bibinfo {author}
  {\bibfnamefont {K.}~\bibnamefont {Araki}}, \bibinfo {author} {\bibfnamefont
  {K.}~\bibnamefont {Katoh}},\ and\ \bibinfo {author} {\bibfnamefont
  {A.}~\bibnamefont {Ochiai}},\ }\bibfield  {title} {\bibinfo {title}
  {{Eigenstate analysis of the crystal electric field at low-symmetry sites:
  Application for an orthogonal site in the tetragonal crystal
  ${\mathrm{Ce}}_{2}{\mathrm{Pd}}_{2}\mathrm{Pb}$}},\ }\href
  {https://doi.org/10.1103/PhysRevB.105.014419} {\bibfield  {journal} {\bibinfo
   {journal} {Phys. Rev. B}\ }\textbf {\bibinfo {volume} {105}},\ \bibinfo
  {pages} {014419} (\bibinfo {year} {2022})}\BibitemShut {NoStop}%
\bibitem [{\citenamefont {Mondal}\ \emph {et~al.}(2018)\citenamefont {Mondal},
  \citenamefont {Bapat}, \citenamefont {Dhar},\ and\ \citenamefont
  {Thamizhavel}}]{CeAgAs2_2018}%
  \BibitemOpen
  \bibfield  {author} {\bibinfo {author} {\bibfnamefont {R.}~\bibnamefont
  {Mondal}}, \bibinfo {author} {\bibfnamefont {R.}~\bibnamefont {Bapat}},
  \bibinfo {author} {\bibfnamefont {S.~K.}\ \bibnamefont {Dhar}},\ and\
  \bibinfo {author} {\bibfnamefont {A.}~\bibnamefont {Thamizhavel}},\
  }\bibfield  {title} {\bibinfo {title} {{Magnetocrystalline anisotropy in the
  Kondo-lattice compound ${\mathrm{CeAgAs}}_{2}$}},\ }\href
  {https://doi.org/10.1103/PhysRevB.98.115160} {\bibfield  {journal} {\bibinfo
  {journal} {Phys. Rev. B}\ }\textbf {\bibinfo {volume} {98}},\ \bibinfo
  {pages} {115160} (\bibinfo {year} {2018})}\BibitemShut {NoStop}%
\bibitem [{\citenamefont {Chang}\ \emph {et~al.}(2018)\citenamefont {Chang},
  \citenamefont {Singh}, \citenamefont {Xu}, \citenamefont {Bian},
  \citenamefont {Huang}, \citenamefont {Hsu}, \citenamefont {Belopolski},
  \citenamefont {Alidoust}, \citenamefont {Sanchez}, \citenamefont {Zheng},
  \citenamefont {Lu}, \citenamefont {Zhang}, \citenamefont {Bian},
  \citenamefont {Chang}, \citenamefont {Jeng}, \citenamefont {Bansil},
  \citenamefont {Hsu}, \citenamefont {Jia}, \citenamefont {Neupert},
  \citenamefont {Lin},\ and\ \citenamefont {Hasan}}]{RAlGe_2018_theory}%
  \BibitemOpen
  \bibfield  {author} {\bibinfo {author} {\bibfnamefont {G.}~\bibnamefont
  {Chang}}, \bibinfo {author} {\bibfnamefont {B.}~\bibnamefont {Singh}},
  \bibinfo {author} {\bibfnamefont {S.-Y.}\ \bibnamefont {Xu}}, \bibinfo
  {author} {\bibfnamefont {G.}~\bibnamefont {Bian}}, \bibinfo {author}
  {\bibfnamefont {S.-M.}\ \bibnamefont {Huang}}, \bibinfo {author}
  {\bibfnamefont {C.-H.}\ \bibnamefont {Hsu}}, \bibinfo {author} {\bibfnamefont
  {I.}~\bibnamefont {Belopolski}}, \bibinfo {author} {\bibfnamefont
  {N.}~\bibnamefont {Alidoust}}, \bibinfo {author} {\bibfnamefont {D.~S.}\
  \bibnamefont {Sanchez}}, \bibinfo {author} {\bibfnamefont {H.}~\bibnamefont
  {Zheng}}, \bibinfo {author} {\bibfnamefont {H.}~\bibnamefont {Lu}}, \bibinfo
  {author} {\bibfnamefont {X.}~\bibnamefont {Zhang}}, \bibinfo {author}
  {\bibfnamefont {Y.}~\bibnamefont {Bian}}, \bibinfo {author} {\bibfnamefont
  {T.-R.}\ \bibnamefont {Chang}}, \bibinfo {author} {\bibfnamefont {H.-T.}\
  \bibnamefont {Jeng}}, \bibinfo {author} {\bibfnamefont {A.}~\bibnamefont
  {Bansil}}, \bibinfo {author} {\bibfnamefont {H.}~\bibnamefont {Hsu}},
  \bibinfo {author} {\bibfnamefont {S.}~\bibnamefont {Jia}}, \bibinfo {author}
  {\bibfnamefont {T.}~\bibnamefont {Neupert}}, \bibinfo {author} {\bibfnamefont
  {H.}~\bibnamefont {Lin}},\ and\ \bibinfo {author} {\bibfnamefont {M.~Z.}\
  \bibnamefont {Hasan}},\ }\bibfield  {title} {\bibinfo {title} {{Magnetic and
  noncentrosymmetric Weyl fermion semimetals in the $\mathit{R}\mathrm{AlGe}$
  family of compounds
  ($\mathit{R}=\mathrm{rare}\phantom{\rule{0.28em}{0ex}}\mathrm{earth}$)}},\
  }\href {https://doi.org/10.1103/PhysRevB.97.041104} {\bibfield  {journal}
  {\bibinfo  {journal} {Phys. Rev. B}\ }\textbf {\bibinfo {volume} {97}},\
  \bibinfo {pages} {041104} (\bibinfo {year} {2018})}\BibitemShut {NoStop}%
\bibitem [{\citenamefont {Puphal}\ \emph {et~al.}(2019)\citenamefont {Puphal},
  \citenamefont {Mielke}, \citenamefont {Kumar}, \citenamefont {Soh},
  \citenamefont {Shang}, \citenamefont {Medarde}, \citenamefont {White},\ and\
  \citenamefont {Pomjakushina}}]{RAlGe_2019}%
  \BibitemOpen
  \bibfield  {author} {\bibinfo {author} {\bibfnamefont {P.}~\bibnamefont
  {Puphal}}, \bibinfo {author} {\bibfnamefont {C.}~\bibnamefont {Mielke}},
  \bibinfo {author} {\bibfnamefont {N.}~\bibnamefont {Kumar}}, \bibinfo
  {author} {\bibfnamefont {Y.}~\bibnamefont {Soh}}, \bibinfo {author}
  {\bibfnamefont {T.}~\bibnamefont {Shang}}, \bibinfo {author} {\bibfnamefont
  {M.}~\bibnamefont {Medarde}}, \bibinfo {author} {\bibfnamefont {J.~S.}\
  \bibnamefont {White}},\ and\ \bibinfo {author} {\bibfnamefont
  {E.}~\bibnamefont {Pomjakushina}},\ }\bibfield  {title} {\bibinfo {title}
  {{Bulk single-crystal growth of the theoretically predicted magnetic Weyl
  semimetals $R\mathrm{AlGe}$ ($R$ = Pr, Ce)}},\ }\href
  {https://doi.org/10.1103/PhysRevMaterials.3.024204} {\bibfield  {journal}
  {\bibinfo  {journal} {Phys. Rev. Mater.}\ }\textbf {\bibinfo {volume} {3}},\
  \bibinfo {pages} {024204} (\bibinfo {year} {2019})}\BibitemShut {NoStop}%
\bibitem [{\citenamefont {Hodovanets}\ \emph {et~al.}(2018)\citenamefont
  {Hodovanets}, \citenamefont {Eckberg}, \citenamefont {Zavalij}, \citenamefont
  {Kim}, \citenamefont {Lin}, \citenamefont {Zic}, \citenamefont {Campbell},
  \citenamefont {Higgins},\ and\ \citenamefont {Paglione}}]{CeAlGe_2018}%
  \BibitemOpen
  \bibfield  {author} {\bibinfo {author} {\bibfnamefont {H.}~\bibnamefont
  {Hodovanets}}, \bibinfo {author} {\bibfnamefont {C.~J.}\ \bibnamefont
  {Eckberg}}, \bibinfo {author} {\bibfnamefont {P.~Y.}\ \bibnamefont
  {Zavalij}}, \bibinfo {author} {\bibfnamefont {H.}~\bibnamefont {Kim}},
  \bibinfo {author} {\bibfnamefont {W.-C.}\ \bibnamefont {Lin}}, \bibinfo
  {author} {\bibfnamefont {M.}~\bibnamefont {Zic}}, \bibinfo {author}
  {\bibfnamefont {D.~J.}\ \bibnamefont {Campbell}}, \bibinfo {author}
  {\bibfnamefont {J.~S.}\ \bibnamefont {Higgins}},\ and\ \bibinfo {author}
  {\bibfnamefont {J.}~\bibnamefont {Paglione}},\ }\bibfield  {title} {\bibinfo
  {title} {{Single-crystal investigation of the proposed type-II Weyl semimetal
  CeAlGe}},\ }\href {https://doi.org/10.1103/PhysRevB.98.245132} {\bibfield
  {journal} {\bibinfo  {journal} {Phys. Rev. B}\ }\textbf {\bibinfo {volume}
  {98}},\ \bibinfo {pages} {245132} (\bibinfo {year} {2018})}\BibitemShut
  {NoStop}%
\bibitem [{\citenamefont {Yashima}\ \emph {et~al.}(1982)\citenamefont
  {Yashima}, \citenamefont {Satoh}, \citenamefont {Mori}, \citenamefont
  {Watanabe},\ and\ \citenamefont {Ohtsuka}}]{CeSi2_1982}%
  \BibitemOpen
  \bibfield  {author} {\bibinfo {author} {\bibfnamefont {H.}~\bibnamefont
  {Yashima}}, \bibinfo {author} {\bibfnamefont {T.}~\bibnamefont {Satoh}},
  \bibinfo {author} {\bibfnamefont {H.}~\bibnamefont {Mori}}, \bibinfo {author}
  {\bibfnamefont {D.}~\bibnamefont {Watanabe}},\ and\ \bibinfo {author}
  {\bibfnamefont {T.}~\bibnamefont {Ohtsuka}},\ }\bibfield  {title} {\bibinfo
  {title} {{Thermal and magnetic properties and crystal structures of
  Ce${\mathrm{Ge}}_{2}$ and Ce${\mathrm{Si}}_{2}$}},\ }\href
  {https://doi.org/https://doi.org/10.1016/0038-1098(82)90237-X} {\bibfield
  {journal} {\bibinfo  {journal} {Solid State Communications}\ }\textbf
  {\bibinfo {volume} {41}},\ \bibinfo {pages} {1} (\bibinfo {year}
  {1982})}\BibitemShut {NoStop}%
\bibitem [{\citenamefont {Mori}\ \emph {et~al.}(1984)\citenamefont {Mori},
  \citenamefont {Sato},\ and\ \citenamefont {Sato}}]{CeSi2-xGa_1984}%
  \BibitemOpen
  \bibfield  {author} {\bibinfo {author} {\bibfnamefont {H.}~\bibnamefont
  {Mori}}, \bibinfo {author} {\bibfnamefont {N.}~\bibnamefont {Sato}},\ and\
  \bibinfo {author} {\bibfnamefont {T.}~\bibnamefont {Sato}},\ }\bibfield
  {title} {\bibinfo {title} {{An electronically-driven volume transition in
  Ce${\mathrm{Si}}_{2-x}$${\mathrm{Ge}}_{x}$}},\ }\href
  {https://doi.org/https://doi.org/10.1016/0038-1098(84)90301-6} {\bibfield
  {journal} {\bibinfo  {journal} {Solid State Communications}\ }\textbf
  {\bibinfo {volume} {49}},\ \bibinfo {pages} {955} (\bibinfo {year}
  {1984})}\BibitemShut {NoStop}%
\bibitem [{\citenamefont {Moshchalkov}\ \emph {et~al.}(1990)\citenamefont
  {Moshchalkov}, \citenamefont {Petrenko},\ and\ \citenamefont
  {Zalyalyutdinov}}]{CeSi2-xGa_1990}%
  \BibitemOpen
  \bibfield  {author} {\bibinfo {author} {\bibfnamefont {V.}~\bibnamefont
  {Moshchalkov}}, \bibinfo {author} {\bibfnamefont {O.}~\bibnamefont
  {Petrenko}},\ and\ \bibinfo {author} {\bibfnamefont {M.}~\bibnamefont
  {Zalyalyutdinov}},\ }\bibfield  {title} {\bibinfo {title} {{The new Kondo
  lattice compounds: Ce${\mathrm{Si}}_{2-x}$${\mathrm{Ge}}_{x}$}},\ }\href
  {https://doi.org/https://doi.org/10.1016/0921-4526(90)90222-G} {\bibfield
  {journal} {\bibinfo  {journal} {Physica B: Condensed Matter}\ }\textbf
  {\bibinfo {volume} {163}},\ \bibinfo {pages} {395} (\bibinfo {year}
  {1990})}\BibitemShut {NoStop}%
\bibitem [{\citenamefont {Synoradzki}\ \emph {et~al.}(2022)\citenamefont
  {Synoradzki}, \citenamefont {Skokowski}, \citenamefont {Koterlyn},
  \citenamefont {Sebesta}, \citenamefont {Legut}, \citenamefont {Toli{\'n}ski}
  \emph {et~al.}}]{CeSi2-xGa_2022}%
  \BibitemOpen
  \bibfield  {author} {\bibinfo {author} {\bibfnamefont {K.}~\bibnamefont
  {Synoradzki}}, \bibinfo {author} {\bibfnamefont {P.}~\bibnamefont
  {Skokowski}}, \bibinfo {author} {\bibfnamefont {M.}~\bibnamefont {Koterlyn}},
  \bibinfo {author} {\bibfnamefont {J.}~\bibnamefont {Sebesta}}, \bibinfo
  {author} {\bibfnamefont {D.}~\bibnamefont {Legut}}, \bibinfo {author}
  {\bibfnamefont {T.}~\bibnamefont {Toli{\'n}ski}}, \emph {et~al.},\ }\bibfield
   {title} {\bibinfo {title} {{Ferromagnetic
  Ce${\mathrm{Si}}_{1.2}$${\mathrm{Ge}}_{0.8}$ alloy: Study on magnetocaloric
  and thermoelectric properties}},\ }\href
  {https://www.sciencedirect.com/science/article/pii/S0304885321010441}
  {\bibfield  {journal} {\bibinfo  {journal} {Journal of Magnetism and Magnetic
  Materials}\ }\textbf {\bibinfo {volume} {547}},\ \bibinfo {pages} {168833}
  (\bibinfo {year} {2022})}\BibitemShut {NoStop}%
\bibitem [{\citenamefont {Dhar}\ \emph {et~al.}(1993)\citenamefont {Dhar},
  \citenamefont {Pattalwar},\ and\ \citenamefont
  {Vijayaraghavan}}]{CeSi2-xGa_1993}%
  \BibitemOpen
  \bibfield  {author} {\bibinfo {author} {\bibfnamefont {S.}~\bibnamefont
  {Dhar}}, \bibinfo {author} {\bibfnamefont {S.}~\bibnamefont {Pattalwar}},\
  and\ \bibinfo {author} {\bibfnamefont {R.}~\bibnamefont {Vijayaraghavan}},\
  }\bibfield  {title} {\bibinfo {title} {{Structural and magnetic properties of
  Ce${\mathrm{Si}}_{2-x}$${\mathrm{Ge}}_{x}$ alloys ($\mathrm{x}$ = 1.5, 1.75
  and 2.0)}},\ }\href
  {https://doi.org/https://doi.org/10.1016/0038-1098(93)90787-N} {\bibfield
  {journal} {\bibinfo  {journal} {Solid State Communications}\ }\textbf
  {\bibinfo {volume} {87}},\ \bibinfo {pages} {409} (\bibinfo {year}
  {1993})}\BibitemShut {NoStop}%
\bibitem [{\citenamefont {Priolkar}\ \emph {et~al.}(1998)\citenamefont
  {Priolkar}, \citenamefont {Prabhu}, \citenamefont {Sarode}, \citenamefont
  {Ganesan}, \citenamefont {Raj},\ and\ \citenamefont
  {Sathyamoorthy}}]{CeSi2-xGa_1998}%
  \BibitemOpen
  \bibfield  {author} {\bibinfo {author} {\bibfnamefont {K.~R.}\ \bibnamefont
  {Priolkar}}, \bibinfo {author} {\bibfnamefont {R.~B.}\ \bibnamefont
  {Prabhu}}, \bibinfo {author} {\bibfnamefont {P.~R.}\ \bibnamefont {Sarode}},
  \bibinfo {author} {\bibfnamefont {V.}~\bibnamefont {Ganesan}}, \bibinfo
  {author} {\bibfnamefont {P.}~\bibnamefont {Raj}},\ and\ \bibinfo {author}
  {\bibfnamefont {A.}~\bibnamefont {Sathyamoorthy}},\ }\bibfield  {title}
  {\bibinfo {title} {{Competition between RKKY and Kondo interactions in
  Ce${\mathrm{Si}}_{2-x}$${\mathrm{Ge}}_{x}$}},\ }\href
  {https://doi.org/10.1088/0953-8984/10/20/009} {\bibfield  {journal} {\bibinfo
   {journal} {Journal of Physics: Condensed Matter}\ }\textbf {\bibinfo
  {volume} {10}},\ \bibinfo {pages} {4413} (\bibinfo {year}
  {1998})}\BibitemShut {NoStop}%
\bibitem [{\citenamefont {Zhang}\ \emph {et~al.}(2024)\citenamefont {Zhang},
  \citenamefont {Dong}, \citenamefont {Bai}, \citenamefont {Liu}, \citenamefont
  {Cheng}, \citenamefont {Li}, \citenamefont {Liu}, \citenamefont {Sun},
  \citenamefont {Huang}, \citenamefont {Ren},\ and\ \citenamefont
  {Chen}}]{CeGaSi_2024_02}%
  \BibitemOpen
  \bibfield  {author} {\bibinfo {author} {\bibfnamefont {L.-B.}\ \bibnamefont
  {Zhang}}, \bibinfo {author} {\bibfnamefont {Q.-X.}\ \bibnamefont {Dong}},
  \bibinfo {author} {\bibfnamefont {J.-L.}\ \bibnamefont {Bai}}, \bibinfo
  {author} {\bibfnamefont {Q.-Y.}\ \bibnamefont {Liu}}, \bibinfo {author}
  {\bibfnamefont {J.-W.}\ \bibnamefont {Cheng}}, \bibinfo {author}
  {\bibfnamefont {C.-D.}\ \bibnamefont {Li}}, \bibinfo {author} {\bibfnamefont
  {P.-Y.}\ \bibnamefont {Liu}}, \bibinfo {author} {\bibfnamefont {Y.-R.}\
  \bibnamefont {Sun}}, \bibinfo {author} {\bibfnamefont {Y.}~\bibnamefont
  {Huang}}, \bibinfo {author} {\bibfnamefont {Z.-A.}\ \bibnamefont {Ren}},\
  and\ \bibinfo {author} {\bibfnamefont {G.-F.}\ \bibnamefont {Chen}},\
  }\bibfield  {title} {\bibinfo {title} {{Magnetism, heat capacity,
  magnetocaloric effect, and magneto-transport properties of heavy fermion
  antiferromagnet CeGaSi}},\ }\href {https://doi.org/10.1088/1674-1056/ad3060}
  {\bibfield  {journal} {\bibinfo  {journal} {Chinese Physics B}\ }\textbf
  {\bibinfo {volume} {33}},\ \bibinfo {pages} {067101} (\bibinfo {year}
  {2024})}\BibitemShut {NoStop}%
\bibitem [{\citenamefont {Bl\"ochl}(1994)}]{paw}%
  \BibitemOpen
  \bibfield  {author} {\bibinfo {author} {\bibfnamefont {P.~E.}\ \bibnamefont
  {Bl\"ochl}},\ }\bibfield  {title} {\bibinfo {title} {Projector augmented-wave
  method},\ }\href {https://doi.org/10.1103/PhysRevB.50.17953} {\bibfield
  {journal} {\bibinfo  {journal} {Phys. Rev. B}\ }\textbf {\bibinfo {volume}
  {50}},\ \bibinfo {pages} {17953} (\bibinfo {year} {1994})}\BibitemShut
  {NoStop}%
\bibitem [{\citenamefont {Kresse}\ and\ \citenamefont
  {Joubert}(1999)}]{vasp_1}%
  \BibitemOpen
  \bibfield  {author} {\bibinfo {author} {\bibfnamefont {G.}~\bibnamefont
  {Kresse}}\ and\ \bibinfo {author} {\bibfnamefont {D.}~\bibnamefont
  {Joubert}},\ }\bibfield  {title} {\bibinfo {title} {From ultrasoft
  pseudopotentials to the projector augmented-wave method},\ }\href
  {https://doi.org/10.1103/PhysRevB.59.1758} {\bibfield  {journal} {\bibinfo
  {journal} {Phys. Rev. B}\ }\textbf {\bibinfo {volume} {59}},\ \bibinfo
  {pages} {1758} (\bibinfo {year} {1999})}\BibitemShut {NoStop}%
\bibitem [{\citenamefont {Kresse}\ and\ \citenamefont
  {Furthm\"uller}(1996)}]{vasp_2}%
  \BibitemOpen
  \bibfield  {author} {\bibinfo {author} {\bibfnamefont {G.}~\bibnamefont
  {Kresse}}\ and\ \bibinfo {author} {\bibfnamefont {J.}~\bibnamefont
  {Furthm\"uller}},\ }\bibfield  {title} {\bibinfo {title} {Efficient iterative
  schemes for \textit{ab} \textit{initio} total-energy calculations using a
  plane-wave basis set},\ }\href {https://doi.org/10.1103/PhysRevB.54.11169}
  {\bibfield  {journal} {\bibinfo  {journal} {Phys. Rev. B}\ }\textbf {\bibinfo
  {volume} {54}},\ \bibinfo {pages} {11169} (\bibinfo {year}
  {1996})}\BibitemShut {NoStop}%
\bibitem [{\citenamefont {Perdew}\ \emph {et~al.}(1996)\citenamefont {Perdew},
  \citenamefont {Burke},\ and\ \citenamefont {Ernzerhof}}]{pbe}%
  \BibitemOpen
  \bibfield  {author} {\bibinfo {author} {\bibfnamefont {J.~P.}\ \bibnamefont
  {Perdew}}, \bibinfo {author} {\bibfnamefont {K.}~\bibnamefont {Burke}},\ and\
  \bibinfo {author} {\bibfnamefont {M.}~\bibnamefont {Ernzerhof}},\ }\bibfield
  {title} {\bibinfo {title} {{Generalized Gradient Approximation Made
  Simple}},\ }\href {https://doi.org/10.1103/PhysRevLett.77.3865} {\bibfield
  {journal} {\bibinfo  {journal} {Phys. Rev. Lett.}\ }\textbf {\bibinfo
  {volume} {77}},\ \bibinfo {pages} {3865} (\bibinfo {year}
  {1996})}\BibitemShut {NoStop}%
\bibitem [{\citenamefont {Dudarev}\ \emph {et~al.}(1998)\citenamefont
  {Dudarev}, \citenamefont {Botton}, \citenamefont {Savrasov}, \citenamefont
  {Humphreys},\ and\ \citenamefont {Sutton}}]{dudarev.botton.98}%
  \BibitemOpen
  \bibfield  {author} {\bibinfo {author} {\bibfnamefont {S.~L.}\ \bibnamefont
  {Dudarev}}, \bibinfo {author} {\bibfnamefont {G.~A.}\ \bibnamefont {Botton}},
  \bibinfo {author} {\bibfnamefont {S.~Y.}\ \bibnamefont {Savrasov}}, \bibinfo
  {author} {\bibfnamefont {C.~J.}\ \bibnamefont {Humphreys}},\ and\ \bibinfo
  {author} {\bibfnamefont {A.~P.}\ \bibnamefont {Sutton}},\ }\bibfield  {title}
  {\bibinfo {title} {Electron-energy-loss spectra and the structural stability
  of nickel oxide: An {LSDA+U} study},\ }\href
  {https://doi.org/10.1103/PhysRevB.57.1505} {\bibfield  {journal} {\bibinfo
  {journal} {Phys. Rev. B}\ }\textbf {\bibinfo {volume} {57}},\ \bibinfo
  {pages} {1505} (\bibinfo {year} {1998})}\BibitemShut {NoStop}%
\bibitem [{\citenamefont {Cho}\ \emph {et~al.}(1996)\citenamefont {Cho},
  \citenamefont {Canfield},\ and\ \citenamefont {Johnston}}]{TbNi2B2C}%
  \BibitemOpen
  \bibfield  {author} {\bibinfo {author} {\bibfnamefont {B.~K.}\ \bibnamefont
  {Cho}}, \bibinfo {author} {\bibfnamefont {P.~C.}\ \bibnamefont {Canfield}},\
  and\ \bibinfo {author} {\bibfnamefont {D.~C.}\ \bibnamefont {Johnston}},\
  }\bibfield  {title} {\bibinfo {title} {{Magnetic anisotropy and weak
  ferromagnetism of single-crystal ${\mathrm{TbNi}}_{2}$${\mathrm{B}}_{2}$C}},\
  }\href {https://doi.org/10.1103/PhysRevB.53.8499} {\bibfield  {journal}
  {\bibinfo  {journal} {Phys. Rev. B}\ }\textbf {\bibinfo {volume} {53}},\
  \bibinfo {pages} {8499} (\bibinfo {year} {1996})}\BibitemShut {NoStop}%
\bibitem [{\citenamefont {Ram}\ \emph {et~al.}(2023{\natexlab{b}})\citenamefont
  {Ram}, \citenamefont {Singh}, \citenamefont {Hooda}, \citenamefont {Singh},
  \citenamefont {Kanchana}, \citenamefont {Kaczorowski},\ and\ \citenamefont
  {Hossain}}]{GdAuGe}%
  \BibitemOpen
  \bibfield  {author} {\bibinfo {author} {\bibfnamefont {D.}~\bibnamefont
  {Ram}}, \bibinfo {author} {\bibfnamefont {J.}~\bibnamefont {Singh}}, \bibinfo
  {author} {\bibfnamefont {M.~K.}\ \bibnamefont {Hooda}}, \bibinfo {author}
  {\bibfnamefont {K.}~\bibnamefont {Singh}}, \bibinfo {author} {\bibfnamefont
  {V.}~\bibnamefont {Kanchana}}, \bibinfo {author} {\bibfnamefont
  {D.}~\bibnamefont {Kaczorowski}},\ and\ \bibinfo {author} {\bibfnamefont
  {Z.}~\bibnamefont {Hossain}},\ }\bibfield  {title} {\bibinfo {title}
  {{Multiple magnetic transitions, metamagnetism, and large magnetoresistance
  in GdAuGe single crystals}},\ }\href
  {https://doi.org/10.1103/PhysRevB.108.235107} {\bibfield  {journal} {\bibinfo
   {journal} {Phys. Rev. B}\ }\textbf {\bibinfo {volume} {108}},\ \bibinfo
  {pages} {235107} (\bibinfo {year} {2023}{\natexlab{b}})}\BibitemShut
  {NoStop}%
\bibitem [{\citenamefont {Jobiliong}\ \emph {et~al.}(2005)\citenamefont
  {Jobiliong}, \citenamefont {Brooks}, \citenamefont {Choi}, \citenamefont
  {Lee},\ and\ \citenamefont {Fisk}}]{CeAgSb2_2005}%
  \BibitemOpen
  \bibfield  {author} {\bibinfo {author} {\bibfnamefont {E.}~\bibnamefont
  {Jobiliong}}, \bibinfo {author} {\bibfnamefont {J.~S.}\ \bibnamefont
  {Brooks}}, \bibinfo {author} {\bibfnamefont {E.~S.}\ \bibnamefont {Choi}},
  \bibinfo {author} {\bibfnamefont {H.}~\bibnamefont {Lee}},\ and\ \bibinfo
  {author} {\bibfnamefont {Z.}~\bibnamefont {Fisk}},\ }\bibfield  {title}
  {\bibinfo {title} {{Magnetization and electrical-transport investigation of
  the dense Kondo system $\mathrm{Ce}\mathrm{Ag}{\mathrm{Sb}}_{2}$}},\ }\href
  {https://doi.org/10.1103/PhysRevB.72.104428} {\bibfield  {journal} {\bibinfo
  {journal} {Phys. Rev. B}\ }\textbf {\bibinfo {volume} {72}},\ \bibinfo
  {pages} {104428} (\bibinfo {year} {2005})}\BibitemShut {NoStop}%
\bibitem [{\citenamefont {Stevens}(1952)}]{CEF_1951}%
  \BibitemOpen
  \bibfield  {author} {\bibinfo {author} {\bibfnamefont {K.~W.~H.}\
  \bibnamefont {Stevens}},\ }\bibfield  {title} {\bibinfo {title} {{Matrix
  Elements and Operator Equivalents Connected with the Magnetic Properties of
  Rare Earth Ions}},\ }\href {https://doi.org/10.1088/0370-1298/65/3/308}
  {\bibfield  {journal} {\bibinfo  {journal} {Proceedings of the Physical
  Society. Section A}\ }\textbf {\bibinfo {volume} {65}},\ \bibinfo {pages}
  {209} (\bibinfo {year} {1952})}\BibitemShut {NoStop}%
\bibitem [{\citenamefont {Hutchings}(1964)}]{CEF_1964}%
  \BibitemOpen
  \bibfield  {author} {\bibinfo {author} {\bibfnamefont {M.~T.}\ \bibnamefont
  {Hutchings}},\ }\bibfield  {title} {\bibinfo {title} {{Point-charge
  calculations of energy levels of magnetic ions in crystalline electric
  fields}},\ }in\ \href
  {https://doi.org/https://doi.org/10.1016/S0081-1947(08)60517-2} {\emph
  {\bibinfo {booktitle} {Solid state physics}}},\ Vol.~\bibinfo {volume} {16}\
  (\bibinfo  {publisher} {Elsevier},\ \bibinfo {year} {1964})\ pp.\ \bibinfo
  {pages} {227--273}\BibitemShut {NoStop}%
\bibitem [{\citenamefont {Zhu}\ \emph {et~al.}(2015)\citenamefont {Zhu},
  \citenamefont {Lu}, \citenamefont {Tong}, \citenamefont {Wang}, \citenamefont
  {Zhou},\ and\ \citenamefont {Zhang}}]{La2ZnIrO6}%
  \BibitemOpen
  \bibfield  {author} {\bibinfo {author} {\bibfnamefont {W.~K.}\ \bibnamefont
  {Zhu}}, \bibinfo {author} {\bibfnamefont {C.-K.}\ \bibnamefont {Lu}},
  \bibinfo {author} {\bibfnamefont {W.}~\bibnamefont {Tong}}, \bibinfo {author}
  {\bibfnamefont {J.~M.}\ \bibnamefont {Wang}}, \bibinfo {author}
  {\bibfnamefont {H.~D.}\ \bibnamefont {Zhou}},\ and\ \bibinfo {author}
  {\bibfnamefont {S.~X.}\ \bibnamefont {Zhang}},\ }\bibfield  {title} {\bibinfo
  {title} {{Strong ferromagnetism induced by canted antiferromagnetic order in
  double perovskite iridates
  $({\mathrm{La}}_{1\ensuremath{-}x}{\mathrm{Sr}}_{x}){}_{2}{\mathrm{ZnIrO}}_{6}$}},\
  }\href {https://doi.org/10.1103/PhysRevB.91.144408} {\bibfield  {journal}
  {\bibinfo  {journal} {Phys. Rev. B}\ }\textbf {\bibinfo {volume} {91}},\
  \bibinfo {pages} {144408} (\bibinfo {year} {2015})}\BibitemShut {NoStop}%
\bibitem [{\citenamefont {Cao}\ \emph {et~al.}(2013)\citenamefont {Cao},
  \citenamefont {Subedi}, \citenamefont {Calder}, \citenamefont {Yan},
  \citenamefont {Yi}, \citenamefont {Gai}, \citenamefont {Poudel},
  \citenamefont {Singh}, \citenamefont {Lumsden}, \citenamefont {Christianson},
  \citenamefont {Sales},\ and\ \citenamefont {Mandrus}}]{La2ZnIrO6_02}%
  \BibitemOpen
  \bibfield  {author} {\bibinfo {author} {\bibfnamefont {G.}~\bibnamefont
  {Cao}}, \bibinfo {author} {\bibfnamefont {A.}~\bibnamefont {Subedi}},
  \bibinfo {author} {\bibfnamefont {S.}~\bibnamefont {Calder}}, \bibinfo
  {author} {\bibfnamefont {J.-Q.}\ \bibnamefont {Yan}}, \bibinfo {author}
  {\bibfnamefont {J.}~\bibnamefont {Yi}}, \bibinfo {author} {\bibfnamefont
  {Z.}~\bibnamefont {Gai}}, \bibinfo {author} {\bibfnamefont {L.}~\bibnamefont
  {Poudel}}, \bibinfo {author} {\bibfnamefont {D.~J.}\ \bibnamefont {Singh}},
  \bibinfo {author} {\bibfnamefont {M.~D.}\ \bibnamefont {Lumsden}}, \bibinfo
  {author} {\bibfnamefont {A.~D.}\ \bibnamefont {Christianson}}, \bibinfo
  {author} {\bibfnamefont {B.~C.}\ \bibnamefont {Sales}},\ and\ \bibinfo
  {author} {\bibfnamefont {D.}~\bibnamefont {Mandrus}},\ }\bibfield  {title}
  {\bibinfo {title} {{Magnetism and electronic structure of
  La${}_{2}$ZnIrO${}_{6}$ and La${}_{2}$MgIrO${}_{6}$: Candidate
  ${J}_{\mathrm{eff}}=\frac{1}{2}$ Mott insulators}},\ }\href
  {https://doi.org/10.1103/PhysRevB.87.155136} {\bibfield  {journal} {\bibinfo
  {journal} {Phys. Rev. B}\ }\textbf {\bibinfo {volume} {87}},\ \bibinfo
  {pages} {155136} (\bibinfo {year} {2013})}\BibitemShut {NoStop}%
\bibitem [{\citenamefont {Wachter}\ \emph {et~al.}(1999)\citenamefont
  {Wachter}, \citenamefont {Degiorgi}, \citenamefont {Wetzel},\ and\
  \citenamefont {Schwer}}]{Ce3Cu3Sb4_01}%
  \BibitemOpen
  \bibfield  {author} {\bibinfo {author} {\bibfnamefont {P.}~\bibnamefont
  {Wachter}}, \bibinfo {author} {\bibfnamefont {L.}~\bibnamefont {Degiorgi}},
  \bibinfo {author} {\bibfnamefont {G.}~\bibnamefont {Wetzel}},\ and\ \bibinfo
  {author} {\bibfnamefont {H.}~\bibnamefont {Schwer}},\ }\bibfield  {title}
  {\bibinfo {title} {{${\mathrm{Ce}}_{3}{\mathrm{Cu}}_{3}{\mathrm{Sb}}_{4}:$ A
  semimetal with a spontaneous magnetic moment}},\ }\href
  {https://doi.org/10.1103/PhysRevB.60.9518} {\bibfield  {journal} {\bibinfo
  {journal} {Phys. Rev. B}\ }\textbf {\bibinfo {volume} {60}},\ \bibinfo
  {pages} {9518} (\bibinfo {year} {1999})}\BibitemShut {NoStop}%
\bibitem [{\citenamefont {Herrmannsdörfer}\ \emph {et~al.}(1999)\citenamefont
  {Herrmannsdörfer}, \citenamefont {Fischer}, \citenamefont {Wachter},
  \citenamefont {Wetzel},\ and\ \citenamefont {Mattenberger}}]{Ce3Cu3Sb4_02}%
  \BibitemOpen
  \bibfield  {author} {\bibinfo {author} {\bibfnamefont {T.}~\bibnamefont
  {Herrmannsdörfer}}, \bibinfo {author} {\bibfnamefont {P.}~\bibnamefont
  {Fischer}}, \bibinfo {author} {\bibfnamefont {P.}~\bibnamefont {Wachter}},
  \bibinfo {author} {\bibfnamefont {G.}~\bibnamefont {Wetzel}},\ and\ \bibinfo
  {author} {\bibfnamefont {K.}~\bibnamefont {Mattenberger}},\ }\bibfield
  {title} {\bibinfo {title} {{Neutron diffraction investigation of magnetic
  ordering in Ce$_3$Cu$_3$Sb$_4$}},\ }\href
  {https://doi.org/https://doi.org/10.1016/S0038-1098(99)00318-X} {\bibfield
  {journal} {\bibinfo  {journal} {Solid State Communications}\ }\textbf
  {\bibinfo {volume} {112}},\ \bibinfo {pages} {135} (\bibinfo {year}
  {1999})}\BibitemShut {NoStop}%
\bibitem [{\citenamefont {Liu}\ \emph {et~al.}(2020)\citenamefont {Liu},
  \citenamefont {Taddei}, \citenamefont {Li}, \citenamefont {Liu},
  \citenamefont {Dhale}, \citenamefont {Kadado}, \citenamefont {Berman},
  \citenamefont {Cruz},\ and\ \citenamefont {Lv}}]{BaFe2Se4}%
  \BibitemOpen
  \bibfield  {author} {\bibinfo {author} {\bibfnamefont {X.}~\bibnamefont
  {Liu}}, \bibinfo {author} {\bibfnamefont {K.~M.}\ \bibnamefont {Taddei}},
  \bibinfo {author} {\bibfnamefont {S.}~\bibnamefont {Li}}, \bibinfo {author}
  {\bibfnamefont {W.}~\bibnamefont {Liu}}, \bibinfo {author} {\bibfnamefont
  {N.}~\bibnamefont {Dhale}}, \bibinfo {author} {\bibfnamefont
  {R.}~\bibnamefont {Kadado}}, \bibinfo {author} {\bibfnamefont
  {D.}~\bibnamefont {Berman}}, \bibinfo {author} {\bibfnamefont {C.~D.}\
  \bibnamefont {Cruz}},\ and\ \bibinfo {author} {\bibfnamefont
  {B.}~\bibnamefont {Lv}},\ }\bibfield  {title} {\bibinfo {title} {{Canted
  antiferromagnetism in the quasi-one-dimensional iron chalcogenide
  ${\mathrm{BaFe}}_{2}{\mathrm{Se}}_{4}$}},\ }\href
  {https://doi.org/10.1103/PhysRevB.102.180403} {\bibfield  {journal} {\bibinfo
   {journal} {Phys. Rev. B}\ }\textbf {\bibinfo {volume} {102}},\ \bibinfo
  {pages} {180403(R)} (\bibinfo {year} {2020})}\BibitemShut {NoStop}%
\bibitem [{\citenamefont {Tan}\ \emph {et~al.}(2024)\citenamefont {Tan},
  \citenamefont {Feng}, \citenamefont {Ma}, \citenamefont {Wang}, \citenamefont
  {Li}, \citenamefont {Wu}, \citenamefont {Huang}, \citenamefont {Lu},\ and\
  \citenamefont {Xiang}}]{TaNi2Te2}%
  \BibitemOpen
  \bibfield  {author} {\bibinfo {author} {\bibfnamefont {H.}~\bibnamefont
  {Tan}}, \bibinfo {author} {\bibfnamefont {Y.}~\bibnamefont {Feng}}, \bibinfo
  {author} {\bibfnamefont {X.}~\bibnamefont {Ma}}, \bibinfo {author}
  {\bibfnamefont {C.}~\bibnamefont {Wang}}, \bibinfo {author} {\bibfnamefont
  {R.}~\bibnamefont {Li}}, \bibinfo {author} {\bibfnamefont {J.}~\bibnamefont
  {Wu}}, \bibinfo {author} {\bibfnamefont {L.}~\bibnamefont {Huang}}, \bibinfo
  {author} {\bibfnamefont {Y.}~\bibnamefont {Lu}},\ and\ \bibinfo {author}
  {\bibfnamefont {B.}~\bibnamefont {Xiang}},\ }\bibfield  {title} {\bibinfo
  {title} {{Canted antiferromagnetism in a van der Waals metallic material
  ${\mathrm{TaNi}}_{2}{\mathrm{Te}}_{2}$}},\ }\href
  {https://doi.org/10.1103/PhysRevMaterials.8.104412} {\bibfield  {journal}
  {\bibinfo  {journal} {Phys. Rev. Mater.}\ }\textbf {\bibinfo {volume} {8}},\
  \bibinfo {pages} {104412} (\bibinfo {year} {2024})}\BibitemShut {NoStop}%
\bibitem [{\citenamefont {Mondal}\ \emph {et~al.}(2022)\citenamefont {Mondal},
  \citenamefont {Mondal}, \citenamefont {Dan}, \citenamefont {Paudyal},
  \citenamefont {Ranganathan},\ and\ \citenamefont {Mazumdar}}]{TbPt3}%
  \BibitemOpen
  \bibfield  {author} {\bibinfo {author} {\bibfnamefont {S.}~\bibnamefont
  {Mondal}}, \bibinfo {author} {\bibfnamefont {B.}~\bibnamefont {Mondal}},
  \bibinfo {author} {\bibfnamefont {S.}~\bibnamefont {Dan}}, \bibinfo {author}
  {\bibfnamefont {D.}~\bibnamefont {Paudyal}}, \bibinfo {author} {\bibfnamefont
  {R.}~\bibnamefont {Ranganathan}},\ and\ \bibinfo {author} {\bibfnamefont
  {C.}~\bibnamefont {Mazumdar}},\ }\bibfield  {title} {\bibinfo {title}
  {{Contrasting magnetic properties of polymorphic TbPt$_3$}},\ }\href
  {https://doi.org/https://doi.org/10.1016/j.jallcom.2022.165942} {\bibfield
  {journal} {\bibinfo  {journal} {Journal of Alloys and Compounds}\ }\textbf
  {\bibinfo {volume} {920}},\ \bibinfo {pages} {165942} (\bibinfo {year}
  {2022})}\BibitemShut {NoStop}%
\bibitem [{\citenamefont {Banda}\ \emph {et~al.}(2018)\citenamefont {Banda},
  \citenamefont {Rai}, \citenamefont {Rosner}, \citenamefont {Morosan},
  \citenamefont {Geibel},\ and\ \citenamefont {Brando}}]{CeIr3Ge7_CEF}%
  \BibitemOpen
  \bibfield  {author} {\bibinfo {author} {\bibfnamefont {J.}~\bibnamefont
  {Banda}}, \bibinfo {author} {\bibfnamefont {B.~K.}\ \bibnamefont {Rai}},
  \bibinfo {author} {\bibfnamefont {H.}~\bibnamefont {Rosner}}, \bibinfo
  {author} {\bibfnamefont {E.}~\bibnamefont {Morosan}}, \bibinfo {author}
  {\bibfnamefont {C.}~\bibnamefont {Geibel}},\ and\ \bibinfo {author}
  {\bibfnamefont {M.}~\bibnamefont {Brando}},\ }\bibfield  {title} {\bibinfo
  {title} {{Crystalline electric field of Ce in trigonal symmetry:
  ${\mathrm{CeIr}}_{3}{\mathrm{Ge}}_{7}$ as a model case}},\ }\href
  {https://doi.org/10.1103/PhysRevB.98.195120} {\bibfield  {journal} {\bibinfo
  {journal} {Phys. Rev. B}\ }\textbf {\bibinfo {volume} {98}},\ \bibinfo
  {pages} {195120} (\bibinfo {year} {2018})}\BibitemShut {NoStop}%
\bibitem [{\citenamefont {Malick}\ \emph {et~al.}(2024)\citenamefont {Malick},
  \citenamefont {Swiatek}, \citenamefont {Bławat}, \citenamefont {Singleton},\
  and\ \citenamefont {Klimczuk}}]{EuAg4Sb2}%
  \BibitemOpen
  \bibfield  {author} {\bibinfo {author} {\bibfnamefont {S.}~\bibnamefont
  {Malick}}, \bibinfo {author} {\bibfnamefont {H.}~\bibnamefont {Swiatek}},
  \bibinfo {author} {\bibfnamefont {J.}~\bibnamefont {Bławat}}, \bibinfo
  {author} {\bibfnamefont {J.}~\bibnamefont {Singleton}},\ and\ \bibinfo
  {author} {\bibfnamefont {T.}~\bibnamefont {Klimczuk}},\ }\bibfield  {title}
  {\bibinfo {title} {{Large magnetoresistance and first-order phase transition
  in antiferromagnetic single-crystalline EuAg$_{4}${Sb}$_{2}$}},\ }\href
  {https://doi.org/10.1103/PhysRevB.110.165149} {\bibfield  {journal} {\bibinfo
   {journal} {Phys. Rev. B}\ }\textbf {\bibinfo {volume} {110}},\ \bibinfo
  {pages} {165149} (\bibinfo {year} {2024})}\BibitemShut {NoStop}%
\bibitem [{EuA(2024)}]{EuAg2Ge2}%
  \BibitemOpen
  \bibfield  {title} {\bibinfo {title} {{Thermodynamic and magnetotransport
  properties of Eu${\mathrm{Ag}}_{2}{\mathrm{Ge}}_{2}$ single crystals}},\
  }\href {https://doi.org/https://doi.org/10.1016/j.jmmm.2023.171517}
  {\bibfield  {journal} {\bibinfo  {journal} {Journal of Magnetism and Magnetic
  Materials}\ }\textbf {\bibinfo {volume} {589}},\ \bibinfo {pages} {171517}
  (\bibinfo {year} {2024})}\BibitemShut {NoStop}%
\bibitem [{\citenamefont {Gopal}(2012)}]{Schottky_CEF}%
  \BibitemOpen
  \bibfield  {author} {\bibinfo {author} {\bibfnamefont {E.}~\bibnamefont
  {Gopal}},\ }\href {https://books.google.co.in/books?id=Rj3jBwAAQBAJ} {\emph
  {\bibinfo {title} {{Specific Heats at Low Temperatures}}}},\ The
  International Cryogenics Monograph Series\ (\bibinfo  {publisher} {Springer
  US},\ \bibinfo {year} {2012})\BibitemShut {NoStop}%
\bibitem [{\citenamefont {Anand}\ and\ \citenamefont
  {Johnston}(2014)}]{EuPd2As2}%
  \BibitemOpen
  \bibfield  {author} {\bibinfo {author} {\bibfnamefont {V.~K.}\ \bibnamefont
  {Anand}}\ and\ \bibinfo {author} {\bibfnamefont {D.~C.}\ \bibnamefont
  {Johnston}},\ }\bibfield  {title} {\bibinfo {title} {{Physical properties of
  Eu${\mathrm{Pd}}_{2}{\mathrm{As}}_{2}$ single crystals}},\ }\href
  {https://doi.org/10.1088/0953-8984/26/28/286002} {\bibfield  {journal}
  {\bibinfo  {journal} {Journal of Physics: Condensed Matter}\ }\textbf
  {\bibinfo {volume} {26}},\ \bibinfo {pages} {286002} (\bibinfo {year}
  {2014})}\BibitemShut {NoStop}%
\bibitem [{\citenamefont {Lyu}\ \emph {et~al.}(2020)\citenamefont {Lyu},
  \citenamefont {Xiang}, \citenamefont {Mi}, \citenamefont {Zhao},
  \citenamefont {Wang}, \citenamefont {Liu}, \citenamefont {Chen},
  \citenamefont {Ren}, \citenamefont {Li},\ and\ \citenamefont
  {Sun}}]{PrAlSi_2020}%
  \BibitemOpen
  \bibfield  {author} {\bibinfo {author} {\bibfnamefont {M.}~\bibnamefont
  {Lyu}}, \bibinfo {author} {\bibfnamefont {J.}~\bibnamefont {Xiang}}, \bibinfo
  {author} {\bibfnamefont {Z.}~\bibnamefont {Mi}}, \bibinfo {author}
  {\bibfnamefont {H.}~\bibnamefont {Zhao}}, \bibinfo {author} {\bibfnamefont
  {Z.}~\bibnamefont {Wang}}, \bibinfo {author} {\bibfnamefont {E.}~\bibnamefont
  {Liu}}, \bibinfo {author} {\bibfnamefont {G.}~\bibnamefont {Chen}}, \bibinfo
  {author} {\bibfnamefont {Z.}~\bibnamefont {Ren}}, \bibinfo {author}
  {\bibfnamefont {G.}~\bibnamefont {Li}},\ and\ \bibinfo {author}
  {\bibfnamefont {P.}~\bibnamefont {Sun}},\ }\bibfield  {title} {\bibinfo
  {title} {{Nonsaturating magnetoresistance, anomalous Hall effect, and
  magnetic quantum oscillations in the ferromagnetic semimetal PrAlSi}},\
  }\href {https://doi.org/10.1103/PhysRevB.102.085143} {\bibfield  {journal}
  {\bibinfo  {journal} {Phys. Rev. B}\ }\textbf {\bibinfo {volume} {102}},\
  \bibinfo {pages} {085143} (\bibinfo {year} {2020})}\BibitemShut {NoStop}%
\bibitem [{\citenamefont {Swami}\ \emph {et~al.}(2025)\citenamefont {Swami},
  \citenamefont {Ram},\ and\ \citenamefont {Hossain}}]{PrGaSi}%
  \BibitemOpen
  \bibfield  {author} {\bibinfo {author} {\bibfnamefont {R.}~\bibnamefont
  {Swami}}, \bibinfo {author} {\bibfnamefont {D.}~\bibnamefont {Ram}},\ and\
  \bibinfo {author} {\bibfnamefont {Z.}~\bibnamefont {Hossain}},\ }\bibfield
  {title} {\bibinfo {title} {{Crystalline electric field and anomalous Hall
  effect in topological semimetal PrGaSi}},\ }\href
  {https://doi.org/https://doi.org/10.1016/j.jmmm.2025.173022} {\bibfield
  {journal} {\bibinfo  {journal} {Journal of Magnetism and Magnetic Materials}\
  }\textbf {\bibinfo {volume} {624}},\ \bibinfo {pages} {173022} (\bibinfo
  {year} {2025})}\BibitemShut {NoStop}%
\bibitem [{\citenamefont {Yang}\ \emph {et~al.}(2020)\citenamefont {Yang},
  \citenamefont {Singh}, \citenamefont {Lu}, \citenamefont {Huang},
  \citenamefont {Bahrami}, \citenamefont {Chiu}, \citenamefont {Graf},
  \citenamefont {Huang}, \citenamefont {Wang}, \citenamefont {Lin},
  \citenamefont {Torchinsky}, \citenamefont {Bansil},\ and\ \citenamefont
  {Tafti}}]{PrAlGe1−xSix_AHE}%
  \BibitemOpen
  \bibfield  {author} {\bibinfo {author} {\bibfnamefont {H.-Y.}\ \bibnamefont
  {Yang}}, \bibinfo {author} {\bibfnamefont {B.}~\bibnamefont {Singh}},
  \bibinfo {author} {\bibfnamefont {B.}~\bibnamefont {Lu}}, \bibinfo {author}
  {\bibfnamefont {C.-Y.}\ \bibnamefont {Huang}}, \bibinfo {author}
  {\bibfnamefont {F.}~\bibnamefont {Bahrami}}, \bibinfo {author} {\bibfnamefont
  {W.-C.}\ \bibnamefont {Chiu}}, \bibinfo {author} {\bibfnamefont
  {D.}~\bibnamefont {Graf}}, \bibinfo {author} {\bibfnamefont {S.-M.}\
  \bibnamefont {Huang}}, \bibinfo {author} {\bibfnamefont {B.}~\bibnamefont
  {Wang}}, \bibinfo {author} {\bibfnamefont {H.}~\bibnamefont {Lin}}, \bibinfo
  {author} {\bibfnamefont {D.}~\bibnamefont {Torchinsky}}, \bibinfo {author}
  {\bibfnamefont {A.}~\bibnamefont {Bansil}},\ and\ \bibinfo {author}
  {\bibfnamefont {F.}~\bibnamefont {Tafti}},\ }\bibfield  {title} {\bibinfo
  {title} {{Transition from intrinsic to extrinsic anomalous Hall effect in the
  ferromagnetic Weyl semimetal PrAl${\mathrm{Ge}}_{1-x}{\mathrm{Si}}_{x}$}},\
  }\href {https://doi.org/10.1063/1.5132958} {\bibfield  {journal} {\bibinfo
  {journal} {APL Materials}\ }\textbf {\bibinfo {volume} {8}},\ \bibinfo
  {pages} {011111} (\bibinfo {year} {2020})}\BibitemShut {NoStop}%
\bibitem [{\citenamefont {Meng}\ \emph {et~al.}(2019)\citenamefont {Meng},
  \citenamefont {Wu}, \citenamefont {Qiu}, \citenamefont {Wang}, \citenamefont
  {Liu}, \citenamefont {Xia}, \citenamefont {Yuan}, \citenamefont {Chang},\
  and\ \citenamefont {Tian}}]{PrAlGe_2019}%
  \BibitemOpen
  \bibfield  {author} {\bibinfo {author} {\bibfnamefont {B.}~\bibnamefont
  {Meng}}, \bibinfo {author} {\bibfnamefont {H.}~\bibnamefont {Wu}}, \bibinfo
  {author} {\bibfnamefont {Y.}~\bibnamefont {Qiu}}, \bibinfo {author}
  {\bibfnamefont {C.}~\bibnamefont {Wang}}, \bibinfo {author} {\bibfnamefont
  {Y.}~\bibnamefont {Liu}}, \bibinfo {author} {\bibfnamefont {Z.}~\bibnamefont
  {Xia}}, \bibinfo {author} {\bibfnamefont {S.}~\bibnamefont {Yuan}}, \bibinfo
  {author} {\bibfnamefont {H.}~\bibnamefont {Chang}},\ and\ \bibinfo {author}
  {\bibfnamefont {Z.}~\bibnamefont {Tian}},\ }\bibfield  {title} {\bibinfo
  {title} {{Large anomalous Hall effect in ferromagnetic Weyl semimetal
  candidate PrAlGe}},\ }\href {https://doi.org/10.1063/1.5090795} {\bibfield
  {journal} {\bibinfo  {journal} {APL Materials}\ }\textbf {\bibinfo {volume}
  {7}},\ \bibinfo {pages} {051110} (\bibinfo {year} {2019})}\BibitemShut
  {NoStop}%
\bibitem [{\citenamefont {Hurd}(2012)}]{AHE}%
  \BibitemOpen
  \bibfield  {author} {\bibinfo {author} {\bibfnamefont {C.}~\bibnamefont
  {Hurd}},\ }\href {https://books.google.co.in/books?id=5aXTBwAAQBAJ} {\emph
  {\bibinfo {title} {{The Hall Effect in Metals and Alloys}}}},\ The
  International Cryogenics Monograph Series\ (\bibinfo  {publisher} {Springer
  US},\ \bibinfo {year} {2012})\BibitemShut {NoStop}%
\bibitem [{\citenamefont {Ram}\ \emph {et~al.}(2023{\natexlab{c}})\citenamefont
  {Ram}, \citenamefont {Singh}, \citenamefont {Hooda}, \citenamefont
  {Pavlosiuk}, \citenamefont {Kanchana}, \citenamefont {Hossain},\ and\
  \citenamefont {Kaczorowski}}]{GdAgGe}%
  \BibitemOpen
  \bibfield  {author} {\bibinfo {author} {\bibfnamefont {D.}~\bibnamefont
  {Ram}}, \bibinfo {author} {\bibfnamefont {J.}~\bibnamefont {Singh}}, \bibinfo
  {author} {\bibfnamefont {M.~K.}\ \bibnamefont {Hooda}}, \bibinfo {author}
  {\bibfnamefont {O.}~\bibnamefont {Pavlosiuk}}, \bibinfo {author}
  {\bibfnamefont {V.}~\bibnamefont {Kanchana}}, \bibinfo {author}
  {\bibfnamefont {Z.}~\bibnamefont {Hossain}},\ and\ \bibinfo {author}
  {\bibfnamefont {D.}~\bibnamefont {Kaczorowski}},\ }\bibfield  {title}
  {\bibinfo {title} {{Electronic structure and physical properties of the
  candidate topological material GdAgGe}},\ }\href
  {https://doi.org/10.1103/PhysRevB.107.085137} {\bibfield  {journal} {\bibinfo
   {journal} {Phys. Rev. B}\ }\textbf {\bibinfo {volume} {107}},\ \bibinfo
  {pages} {085137} (\bibinfo {year} {2023}{\natexlab{c}})}\BibitemShut
  {NoStop}%
\bibitem [{\citenamefont {Hundley}\ \emph {et~al.}(1997)\citenamefont
  {Hundley}, \citenamefont {Thompson}, \citenamefont {Canfield},\ and\
  \citenamefont {Fisk}}]{YbPtBi}%
  \BibitemOpen
  \bibfield  {author} {\bibinfo {author} {\bibfnamefont {M.~F.}\ \bibnamefont
  {Hundley}}, \bibinfo {author} {\bibfnamefont {J.~D.}\ \bibnamefont
  {Thompson}}, \bibinfo {author} {\bibfnamefont {P.~C.}\ \bibnamefont
  {Canfield}},\ and\ \bibinfo {author} {\bibfnamefont {Z.}~\bibnamefont
  {Fisk}},\ }\bibfield  {title} {\bibinfo {title} {{Electronic transport
  properties of the semimetallic heavy fermion YbBiPt}},\ }\href
  {https://doi.org/10.1103/PhysRevB.56.8098} {\bibfield  {journal} {\bibinfo
  {journal} {Phys. Rev. B}\ }\textbf {\bibinfo {volume} {56}},\ \bibinfo
  {pages} {8098} (\bibinfo {year} {1997})}\BibitemShut {NoStop}%
\bibitem [{\citenamefont {Cho}\ \emph {et~al.}(2023)\citenamefont {Cho},
  \citenamefont {Shon}, \citenamefont {Kyoo}, \citenamefont {Bae},
  \citenamefont {Lee}, \citenamefont {Yun}, \citenamefont {Yoon}, \citenamefont
  {Cho}, \citenamefont {Rawat}, \citenamefont {Kim},\ and\ \citenamefont
  {Rhyee}}]{NdAlGe_AHE}%
  \BibitemOpen
  \bibfield  {author} {\bibinfo {author} {\bibfnamefont {K.}~\bibnamefont
  {Cho}}, \bibinfo {author} {\bibfnamefont {W.~H.}\ \bibnamefont {Shon}},
  \bibinfo {author} {\bibfnamefont {K.}~\bibnamefont {Kyoo}}, \bibinfo {author}
  {\bibfnamefont {J.}~\bibnamefont {Bae}}, \bibinfo {author} {\bibfnamefont
  {J.}~\bibnamefont {Lee}}, \bibinfo {author} {\bibfnamefont {J.~H.}\
  \bibnamefont {Yun}}, \bibinfo {author} {\bibfnamefont {S.}~\bibnamefont
  {Yoon}}, \bibinfo {author} {\bibfnamefont {B.}~\bibnamefont {Cho}}, \bibinfo
  {author} {\bibfnamefont {P.}~\bibnamefont {Rawat}}, \bibinfo {author}
  {\bibfnamefont {Y.-K.}\ \bibnamefont {Kim}},\ and\ \bibinfo {author}
  {\bibfnamefont {J.-S.}\ \bibnamefont {Rhyee}},\ }\bibfield  {title} {\bibinfo
  {title} {{Large anomalous Hall effect and intrinsic Berry curvature in
  magnetic Weyl semimetal NdAlGe}},\ }\href
  {https://doi.org/https://doi.org/10.1016/j.mtcomm.2023.106411} {\bibfield
  {journal} {\bibinfo  {journal} {Materials Today Communications}\ }\textbf
  {\bibinfo {volume} {35}},\ \bibinfo {pages} {106411} (\bibinfo {year}
  {2023})}\BibitemShut {NoStop}%
\bibitem [{\citenamefont {Tian}\ \emph {et~al.}(2009)\citenamefont {Tian},
  \citenamefont {Ye},\ and\ \citenamefont {Jin}}]{Scaling_AHE_01}%
  \BibitemOpen
  \bibfield  {author} {\bibinfo {author} {\bibfnamefont {Y.}~\bibnamefont
  {Tian}}, \bibinfo {author} {\bibfnamefont {L.}~\bibnamefont {Ye}},\ and\
  \bibinfo {author} {\bibfnamefont {X.}~\bibnamefont {Jin}},\ }\bibfield
  {title} {\bibinfo {title} {{Proper Scaling of the Anomalous Hall Effect}},\
  }\href {https://doi.org/10.1103/PhysRevLett.103.087206} {\bibfield  {journal}
  {\bibinfo  {journal} {Phys. Rev. Lett.}\ }\textbf {\bibinfo {volume} {103}},\
  \bibinfo {pages} {087206} (\bibinfo {year} {2009})}\BibitemShut {NoStop}%
\bibitem [{\citenamefont {Wu}\ \emph {et~al.}(2013)\citenamefont {Wu},
  \citenamefont {Li}, \citenamefont {Xu}, \citenamefont {Hou},\ and\
  \citenamefont {Jin}}]{Scaling_AHE_02}%
  \BibitemOpen
  \bibfield  {author} {\bibinfo {author} {\bibfnamefont {L.}~\bibnamefont
  {Wu}}, \bibinfo {author} {\bibfnamefont {Y.}~\bibnamefont {Li}}, \bibinfo
  {author} {\bibfnamefont {J.}~\bibnamefont {Xu}}, \bibinfo {author}
  {\bibfnamefont {D.}~\bibnamefont {Hou}},\ and\ \bibinfo {author}
  {\bibfnamefont {X.}~\bibnamefont {Jin}},\ }\bibfield  {title} {\bibinfo
  {title} {{Anisotropic intrinsic anomalous Hall effect in epitaxial Fe films
  on GaAs(111)}},\ }\href {https://doi.org/10.1103/PhysRevB.87.155307}
  {\bibfield  {journal} {\bibinfo  {journal} {Phys. Rev. B}\ }\textbf {\bibinfo
  {volume} {87}},\ \bibinfo {pages} {155307} (\bibinfo {year}
  {2013})}\BibitemShut {NoStop}%
\end{thebibliography}%

\end{document}